\documentclass{aa}

\usepackage{graphicx}
\usepackage{subcaption}
\usepackage{amssymb}
\usepackage{xcolor}
\usepackage{placeins}
\usepackage{booktabs}
\usepackage{orcidlink}
\usepackage{placeins}

\newcommand{\cii}{[C\,{\sc ii}]}
\newcommand{\ci}{[C\,{\sc i}]}
\newcommand{\hi}{H\,{\sc i}}
\newcommand{\nii}{[N\,{\sc ii}]}

\usepackage{txfonts}

\begin{document}

   \title{Tracing the Hidden Molecular Gas in the Small Magellanic Cloud with SOFIA and APEX}
   \titlerunning{Tracing the Hidden Molecular Gas in the Small Magellanic Cloud with SOFIA and APEX}

   \author{K. Rodriguez-Ambler
          \inst{1,2}\orcidlink{0009-0009-5434-2050}
          , R. Herrera-Camus
\inst{1,2,3}\orcidlink{0000-0002-2775-0595}, M. Rúbio\inst{4}\orcidlink{0000-0002-5307-5941
}, A. D. Bolatto\inst{5}\orcidlink{0000-0002-5480-5686}, D. Riquelme-V\'asquez\inst{6,7}\orcidlink{0000-0001-5389-0535}, K. Jameson \inst{8}\orcidlink{0000-0001-7105-0994}, C. Muñoz-López\inst{9}\orcidlink{0000-0002-7255-3806
},  E. J. Tarantino \inst{10}\orcidlink{0000-0003-1356-1096} \and J. Stutzki \inst{11}\orcidlink{0000-0001-7658-4397}
          }

   \institute{Departamento de Astronomía, Universidad de Concepción,
              Barrio Universitario, Concepción, Chile
         \and
         Millennium Nucleus for Galaxies (MINGAL), Concepción, Chile
        \and
        Max-Planck-Institut f\"ur extraterrestrische Physik, Giessenbachstrasse 1, 85748 Garching, Germany
        \and
             Departamento de Astronomía, Universidad de Chile, Camino El Observatorio 1515, Las Condes, Santiago, Chile
        \and
        Department of Astronomy and Joint Space-Science Institute, University of Maryland, 4296 Stadium Drive, College Park, MD 20742, USA
        \and
        Max-Planck-Institut f\"ur Radioastronomie, Auf dem H\"ugel 69, 53121 Bonn, Germany
        \and
        Departamento de Astronomía, Universidad de La Serena, Raúl Bitrán 1305, la Serena, Chile
        \and
        Caltech Owens Valley Radio Observatory, Pasadena, CA 91125, USA
        \and
        Leibniz-Institut für Astrophysik Potsdam (AIP), An der Sternwarte 16, 14482, Potsdam, Germany
         \and
        Space Telescope Science Institute, 3700 San Martin Drive, Baltimore, MD 21218, USA
        \and
        I. Physikalisches Institut, Universität zu Köln, Zülpicher Str. 77, D-50937 Köln, Germany}

   \date{Received XXXX, XXXX; accepted XXXX, XXXX}

    \abstract
    {In low-metallicity environments, the reduced dust abundance debilitates the shielding of molecular clouds against far-ultraviolet (FUV) radiation, enhancing CO photodissociation and producing extended envelopes of “CO-faint” or “CO-dark” molecular gas. The \cii\ 158 $\mu$m fine-structure line is one of the brightest coolants of the interstellar medium (ISM) and a key tracer of these CO-faint molecular phases.}
    {We aim to characterize the multiphase origin of the \cii\ line emission and quantify the fraction of molecular gas that remains undetected in CO across several regions of the Small Magellanic Cloud (SMC), the nearest low-metallicity ($\sim$0.2 $Z_\odot$) galaxy that provides an ideal laboratory for testing photodissociation region (PDR) models under conditions that resemble those in early galaxies.}
    {We combine high spectral resolution SOFIA/GREAT and upGREAT \cii\ 158 $\mu$m observations with APEX CO(2–1) data and the { ASKAP+Parkes} H I 21 cm map to perform a spectral decomposition of the \cii\ emission into its atomic and molecular contributions. Using the linear decomposition approach, we derive the \cii\ fractions associated with H I and H$_2$, and convert the molecular component into column densities assuming typical PDR conditions constrained by previous SMC studies. We then compare the resulting molecular gas content with CO-based estimates and with the local star formation rate surface density derived from H$\alpha$ and $L_{\mathrm{TIR}}$ tracers.}
     {The \cii\ line width in the analyzed SMC lines of sight consistently falls between the H I and CO line widths, indicating that the \cii\ emission arises from both molecular and atomic gas. On average, $(  {78} \pm 1)\%$ of the \cii\ intensity originates from the molecular phase, while $(  {18} \pm 1)\%$ is associated with atomic gas. The \cii-traced H$_2$ accounts for approximately $(77 \pm 4)\%$ of the total molecular gas column density, implying that the SMC’s molecular reservoir is dominated by CO-faint gas. This estimation yields a H$_2$-to-CO conversion factor on 4~pc scales X$_{\rm CO}\approx  {8.9}\times10^{20}$ cm$^{-2}$(K km s$^{-1}$)$^{-1}$ in CO-emitting regions,  {4.5} times larger than the canonical Galactic conversion factor (X$_{\rm CO,MW}\approx2\times10^{20}$ cm$^{-2}$(K km s$^{-1}$)$^{-1}$). We find no significant correlation between the CO-dark fraction and either total column density or $\Sigma_{\mathrm{SFR}}$, suggesting that the abundance of CO-faint gas is primarily governed by the efficiency of dust and gas shielding rather than by current star formation activity.}
    {Our results demonstrate that \cii\ is the dominant tracer of molecular gas in the SMC and confirm the prevalence of an extended, stable CO-faint molecular phase in low-metallicity environments. This has important implications for interpreting \cii\ observations in unresolved galaxies, both locally and at high redshift, where CO may not fully trace the molecular gas reservoir.}
    \titlerunning{Tracing the Hidden Molecular Gas in the Small Magellanic Cloud with SOFIA and APEX.}
    \authorrunning{Rodriguez-Ambler et al.}
    
   \keywords{ ISM: general -- ISM: photon-dominated region (PDR) -- ISM: Small Magellanic Cloud -- Galaxies: ISM}

  \maketitle

\section{Introduction}
Star formation is regulated by the availability, distribution, and physical conditions of the cold interstellar medium (ISM). In metal-poor environments, the reduced dust abundance weakens the shielding against ultraviolet (UV) photons, altering the chemical and thermal balance between the atomic and molecular phases, as well as the observability of the molecular gas. As a consequence, CO molecules are more easily photodissociated, while H$_2$ remains self-shielded. This leads to extended envelopes of CO-faint molecular gas, commonly referred to as CO-dark H$_2$, that are invisible in CO rotational transitions but still contribute substantially to the total molecular reservoir \citep[e.g.,][]{Wolfire2010, Bolatto2013}.

A powerful tracer of this molecular gas is the \cii\ 158~$\mu$m fine-structure line, one of the brightest cooling lines of the ISM \citep[e.g.,][]{Wolfire2003}. The \cii\ transition arises from the ${}^2P_{3/2}\rightarrow{}^2P_{1/2}$ fine-structure splitting of singly ionized carbon (C$^+$). 
Because the ionization potential of carbon (11.26~eV) is slightly below that of hydrogen (13.6~eV), C$^+$ can exist in a variety of environments, ranging from photodissociation regions and the cold and warm neutral medium (CNM/WNM) to partially ionized gas. This makes \cii\ a key multiphase coolant of the ISM \citep[e.g.,][]{Pineda2013,Langer2014,Croxall2017,Pineda2017}. Moreover, previous studies have demonstrated a strong correlation between \cii\ emission and star formation rate (SFR) surface density \cite[e.g.,][]{delooze2014, HerreraCamus2015, HerreraCamus2018}.

   \begin{figure*}
   \centering
   \includegraphics[width=18cm]{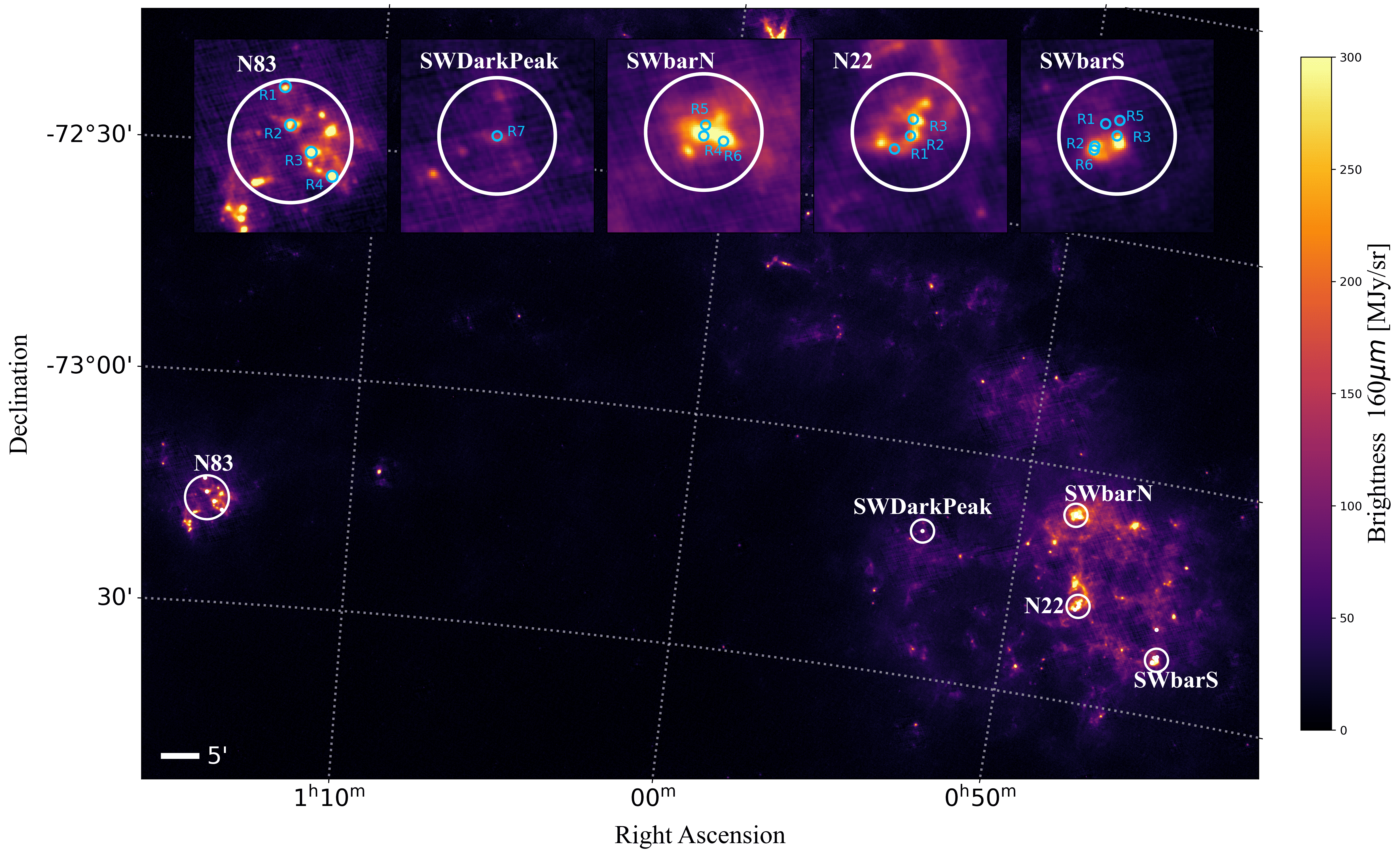}
   \caption{Far-infrared dust emission in the Small Magellanic Cloud. The background image shows the $160~\mu$m surface-brightness map obtained with the \textit{Herschel}/PACS instrument as part of the HERITAGE survey. White circles indicate the SMC regions analyzed in this study, while light-blue circles mark the observed regions listed in Table~\ref{regions table}.} \label{SMC}
    \end{figure*}

In the PDR framework developed by \citep{Kaufman1999, Kaufman2006}, \cii\ emission arises mainly from the surface layers of molecular clouds exposed to far-ultraviolet radiation, where hydrogen transitions from the atomic to the molecular phase and carbon changes from C$^+$ to C and then to CO. The relative intensities of \cii, \ci, and CO therefore describe the structure of the gas across this transition. In dense PDRs, \cii\ traces the warm molecular surface where H$_2$ is already formed but CO is photodissociated, while in diffuse gas it can originate from neutral atomic layers or low-ionization regions. Interpreting \cii\ as a molecular tracer thus requires separating its various phase contributions, ideally by comparing velocity-resolved spectra of \cii, CO, and \hi, and by constraining possible ionized contributions through lines such as [N \textsc{ii}] 205~$\mu$m \citep{HerreraCamus2016, Tarantino2021}.

At low metallicity, where both the dust-to-gas ratio and carbon abundance are reduced, PDR models predict an increase in the \cii-emitting envelope relative to the CO core (e.g. \citealp{Wolfire2010, Hollenbach2012}). This shift enhances the \cii/CO luminosity ratio and implies that a large fraction of the molecular gas is CO-faint. Empirically, this behavior has been observed in nearby dwarf galaxies such as the LMC and SMC \citep{IsraelMaloney2011, Jameson2018, Okada2019}, as well as in other low-metallicity systems \citep{Cormier2019, Madden2020}.

The Small Magellanic Cloud provides an ideal laboratory for studying this effect. Its proximity ($\sim$61~kpc), stellar mass (log($M_*/M_\odot)=8.5$; \citealt{Skibba2012}), and metallicity of 0.2~$Z_\odot$ \citep{Kurt1999, Pagel2003} allow spatially resolved studies of the ISM at sub-parsec scales while sampling a regime comparable to that of early galaxies. Previous works have revealed CO-faint molecular gas on large scales in the SMC (e.g. \citealp{Bolatto2011}), and velocity-resolved SOFIA/GREAT maps for four SMC regions (N66, N25+N26, and N88) have shown that a large fraction ($\sim$60\%) of the molecular gas resides in extended CO-faint envelopes traced by \cii\ emission, highlighting the role of PDR physics in shaping molecular cloud structure at low metallicity \citep{RequenaTorres2016}. Expanding on this framework, \citet{Pineda2017} mapped the transition from atomic to molecular gas across the Magellanic Clouds, establishing that while dense PDR envelopes globally dominate the \cii\ emission, the local contribution from  diffuse CO-dark H$_2$ clouds varies drastically (from 10\% to 80\%) depending on the radiation field. A systematic kinematic decomposition of \cii\ into its atomic and molecular components across multiple regions, and a quantitative census of the molecular gas reservoir traced by \cii\ at high velocity resolution, remain essential for fully constraining the phase balance of carbon and the true molecular gas budget in the SMC.  \citet{Jameson2018} expanded the investigation of the relationship between \cii\ -bright and CO-bright gas in the SMC using \textit{Herschel} and ALMA spectroscopy of five star-forming, extended regions: SWbarS \citep[containing the cloud SMC-B1 and the nebula N13,][]{Rubio1993}, SWbarN (containing the nebula N27/LIRS 49), N22, N83 and SWDarkPeak, suggesting that a $\sim$70\% of the molecular gas is coming from \cii\ -bright regions on average. By comparing integrated intensities and spatial distributions, they showed that \cii\ preferentially traces extended molecular envelopes surrounding compact CO cores, consistent with PDR expectations at low-metallicity environments.

  \begin{table*}[!t] 
\caption{Observed SMC regions with the observed (filled squares) and unobserved (open squares) transitions  and their coordinates (Eq.(J2000.0)), radial velocity, and different molecular, atomic, and ionic transitions that are available and studied in this work.}             
\label{regions table}      
\centering          
\begin{tabular}{c c c c c c c c c c c}    
\toprule\toprule 
Region & RA & Dec & $v_{\rm LSR}$\:[km s$^{-1}$] & \hi & \cii & \ci & $^{12}$CO(2-1) & $^{12}$CO(3-2) & $^{13}$CO(2-1)& $^{13}$CO(3-2) \\ 
\hline       

   N83-R1 & 1:14:24 & -73:14:00  & 165 & $\blacksquare$ & $\blacksquare$ & $\blacksquare$ & $\blacksquare$ & $\blacksquare$ & $\square$ & $\square$\\  
   N83-R2 & 1:14:18 & -73:15:45    & 165 & $\blacksquare$ & $\blacksquare$ & $\blacksquare$ & $\blacksquare$ & $\blacksquare$ & $\square$ & $\square$\\
   N83-R3 & 1:14:04 & -73:16:57    & 165 & $\blacksquare$ & $\blacksquare$ & $\blacksquare$ & $\blacksquare$ & $\blacksquare$ & $\blacksquare$ & $\square$ \\
   N83-R4 & 1:13:50 & -73:18:00   & 165 & $\blacksquare$ & $\blacksquare$ & $\blacksquare$ & $\blacksquare$ & $\blacksquare$ & $\square$ & $\square$\\  
   SWDarkPeak & 0:52:53 & -73:11:06  & 155 &  $\blacksquare$ & $\blacksquare$ & $\blacksquare$ & $\blacksquare$ & $\square$ & $\blacksquare$ & $\square$\\
   SWbarN-R4 & 0:48:27	& -73:06:02   & 115 & $\blacksquare$ & $\blacksquare$ & $\blacksquare$ & $\blacksquare$ & $\square$ & $\square$ & $\blacksquare$  \\
   SWbarN-R5 & 0:48:20	& -73:05:45  & 115 &  $\blacksquare$ & $\blacksquare$ & $\blacksquare$ & $\blacksquare$ & $\square$ & $\blacksquare$ & $\blacksquare$\\
   SWbarN-R6 & 0:48:18 & -73:06:05  & 115 &  $\blacksquare$ & $\blacksquare$ & $\blacksquare$ & $\blacksquare$ & $\blacksquare$ & $\square$ & $\blacksquare$\\
   N22-R1 & 0:47:59 & -73:18:01   & 120 &  $\blacksquare$ & $\blacksquare$ & $\blacksquare$ & $\blacksquare$ & $\square$ & $\blacksquare$ & $\blacksquare$\\  
   N22-R2 & 0:47:54 & -73:17:37    & 120 & $\blacksquare$ & $\blacksquare$ & $\blacksquare$ & $\blacksquare$ & $\square$ & $\square$ & $\blacksquare$\\
   N22-R3 & 0:47:54 & -73:17:11    & 120 &  $\blacksquare$ & $\blacksquare$ & $\blacksquare$ & $\blacksquare$  & $\square$ & $\square$ & $\blacksquare$\\
   SWbarS-R1 & 0:45:23	& -73:22:22  & 120 &  $\blacksquare$ &$\square$ & $\blacksquare$ & $\blacksquare$ & $\blacksquare$ & $\square$ & $\square$\\
   SWbarS-R2 & 0:45:26	& -73:22:59   & 120 &  $\blacksquare$ &$\square$ & $\blacksquare$ & $\blacksquare$ & $\blacksquare$ & $\square$ & $\square$\\
   SWbarS-R3 & 0:45:19 & -73:22:40   & 120 &  $\blacksquare$ &$\square$ & $\blacksquare$ & $\blacksquare$ & $\blacksquare$ &  $\blacksquare$ & $\square$\\
   SWbarS-R4 & 0:45:29 & -73:18:45   & 120 & $\blacksquare$  &$\square$ & $\blacksquare$ & $\blacksquare$  & $\blacksquare$  & $\blacksquare$ & $\blacksquare$\\
   SWbarS-R5 & 0:45:18 & -73:22:13  & 130 &  $\blacksquare$ & $\blacksquare$ & $\blacksquare$ & $\blacksquare$  & $\square$ & $\blacksquare$ & $\square$\\
   SWbarS-R6 & 0:45:27 & -73:23:05  & 130 &  $\blacksquare$ & $\blacksquare$ & $\blacksquare$ & $\blacksquare$ & $\square$ & $\square$ & $\square$\\

\hline                  
\end{tabular}
\end{table*}

In this work, we combine new velocity-resolved SOFIA/GREAT and upGREAT \cii\ 
158~$\mu$m observations towards 10 lines of sight across the SMC with APEX in CO, and ASKAP+Parkes in \hi\ data  to (i) 
determine the kinematic properties (centroid and linewidth) of \cii\ relative to atomic and molecular gas, (ii) decompose the \cii\ emission into its molecular and atomic components, (iii) estimate the H$_2$ column densities traced by \cii\ and compare them with CO-based values, and (iv) relate these results to the SMC’s dust and star-formation properties derived from infrared SED modeling. Together, these analysis provide a self-consistent picture of how \cii\ emission traces the multiphase ISM in a low-metallicity environment and quantify the fraction of the molecular gas reservoir that remains undetected in CO.

The paper is organized as follows. In Section \ref{observations}, we describe the observations used for the \cii\ analysis and the ancillary data employed to constrain the contributions of the different gas phases to the \cii\ 158~$\mu$m emission. Section~\ref{LTIR and SFR} presents the computation of the obscured and unobscured star formation rates. We report the methodology and the results of the \cii\ emission decomposition in Section \ref{results}. The estimation and analysis of the total molecular gas column density estimation, and the correlation between the CO-dark gas fractions and the star formation rates across the SMC regions can be found in Section \ref{CO faint gas}. Then, in Section \ref{darkpeak}, we look into the nature of the SWDarkPeak region to study the structure and transition between CO-bright and CO-dark gas. Finally, in Section \ref{conclusions} we make our final remarks on the results.

\section{Observations}\label{observations}

We observed  17 line-of-sight (LOS) with SOFIA  distributed in five SMC regions, namely N83, SWDarkPeak, SWbarN, N22, and SWbarS. These pointings are overlaid on the 160$\mu m$ emission map of the SMC, as shown in Figure~\ref{SMC}, and sample a diverse range of environments in terms of star formation activity, radiation field strength, and gas content. In this study, we make use of existing  CO and \hi\ observations.

\subsection{SOFIA \cii\ 158~$\mu$m}

Cycle~3 SOFIA \cii\ ($^2$P$_{3/2}-^2$P$_{1/2}$) at 1900.5369 GHz (rest frame) observations ($\lambda\approx157.7$~$\mu$m) were obtained for six regions in the SMC (Proposal ID 03\_0120, PI Herrera-Camus; N22-R1, N22-R2, N22-R3, SWbarN-R4, SWbarN-R5, and SWbarN-R6) using the German Receiver for Astronomy at Terahertz Frequencies (GREAT). The \cii\ line was detected with the L2 channel, which covers a radio-frequency range of 1815-1910~GHz. GREAT is a dual-channel heterodyne instrument for far-infrared (FIR) spectroscopy that provides high spectral resolution in several frequency windows in the 0.490–4.747~THz range, with a beam size of 14.1\arcsec\ \citep{Heyminck2012}, which corresponds to $\sim4.3$~pc for the SMC. The velocity resolution is $\Delta V\sim0.1$ km s$^{-1}$.

Cycle~5 SOFIA \cii\ 158~$\mu$m observations targeting seven additional SMC regions (Proposal ID 05\_0210, PI Herrera-Camus; N83-R1, N83-R2, N83-R3, N83-4, SWDarkPeak-R7, SWbarS-R5, and SWbarS-R6) were obtained with the upgraded GREAT (upGREAT) instrument, and a velocity resolution $\Delta V=0.4$ km s$^{-1}$. The Low Frequency Array receiver (LFA) was used to detect the \cii\ 158~$\mu$m transition over the frequency range 1459.6–1900.5~GHz. upGREAT is a multi-pixel, dual-polarization heterodyne receiver for FIR spectroscopy, providing high spectral resolution ($  {R \sim 10^7}$) over multiple frequency bands in the 1.83–2.07~THz range for the LFA. The instrument’s hexagonal array configuration provides seven beams per polarization, each with a FWHM beam size of $\sim$14.1\arcsec\ \citep{upGREAT} ($\sim4.3$~pc). 

\begin{figure*}[!t]
   \centering
    \includegraphics[width=1\linewidth]{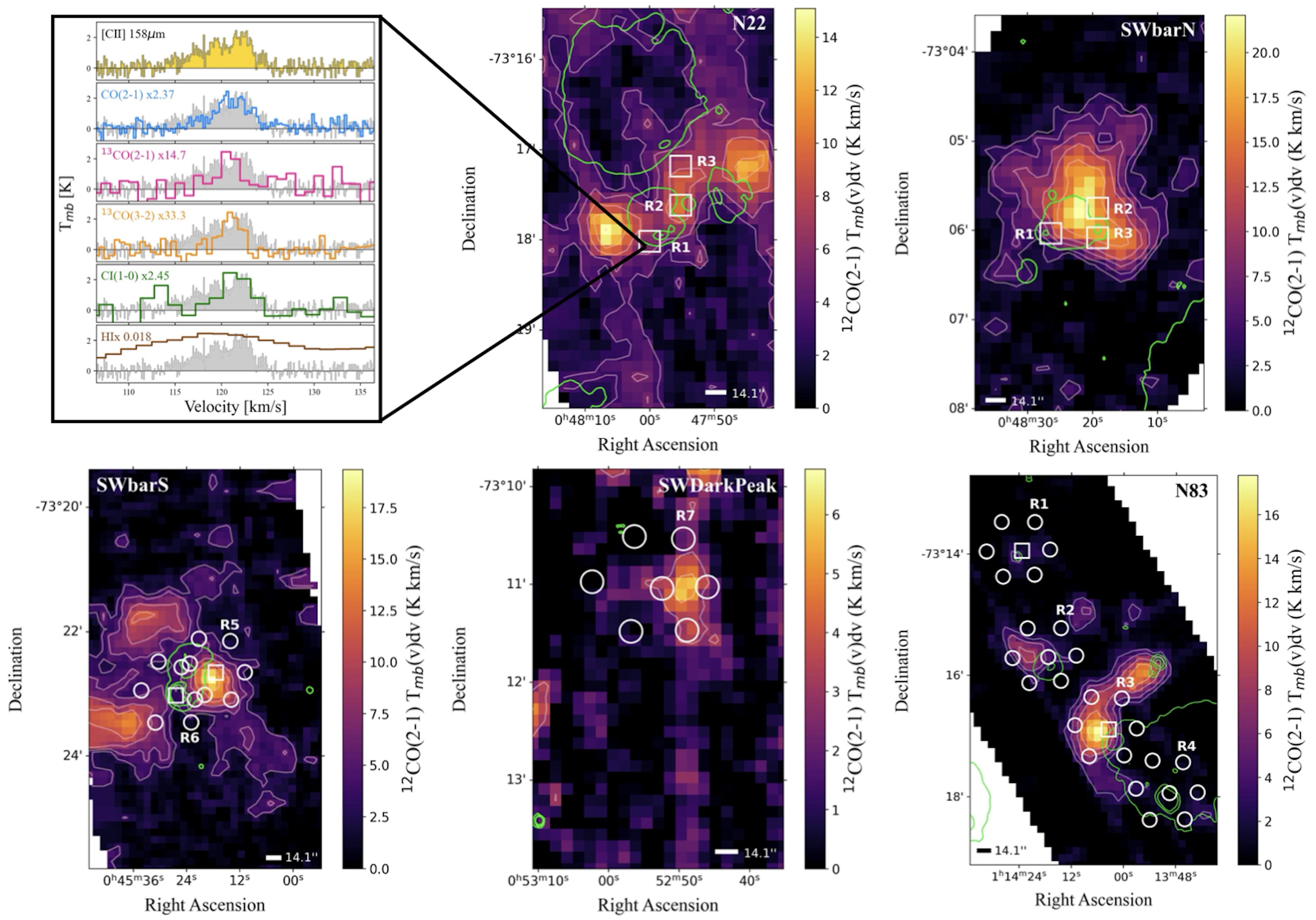}
    \caption{Spectral emission lines of the SMC region N22-R1 in \cii, \hi, \ci, and the different CO transitions (left panel), normalized to the peak intensity of the \cii\ line. The \cii\ 158~$\mu$m line profile lies between the \hi\ and the low-$J$ CO line profiles. Integrated intensity maps of APEX $^{12}$CO(2--1) emission for the SMC regions N22 and SWbarN (top right), and SWbarS, SWDarkPeak, and N83 (bottom panels). Light-pink contours show $^{12}$CO(2--1) emission at 3$\sigma$, 5$\sigma$, 8$\sigma$, 11$\sigma$, and 15$\sigma$ levels in K km s$^{-1}$, while lime-green contours trace H$\alpha$ emission at 3$\sigma$, 5$\sigma$, and 10$\sigma$ levels in erg cm$^{-2}$ s$^{-1}$. White circles indicate the locations of the 14.1\arcsec\ SOFIA/GREAT (top panels) and upGREAT 7-pixel array (bottom panels) 158~$\mu$m observations for each SMC region, labeled as R$_i$, with $i=1$--7. White squares indicate the regions selected  {for the "CO-dark" gas fraction estimation} (see Sect.~\ref{sample selection}).}

    \label{SMC regions}
   \end{figure*}

The targets from Cycle 3 and Cycle 5 are shown in Figure \ref{SMC regions}, overlaid on the CO(2–1) integrated intensity maps. The selected SOFIA/GREAT pointings are distributed across the brightest CO structures in each region. In N22 and SWbarN (top panels), the observed positions trace the main molecular ridge and its associated star-forming clumps. In SWbarS and N83 (bottom left and right panels), the upGREAT array covers compact CO condensations as well as adjacent lower-intensity regions. The H$\alpha$ contours reveal the location of ionized gas relative to the molecular structures, showing that 9 of the \cii\ pointings lie at the interface between bright CO emission and H$\alpha$ peaks, consistent with photodissociation region interfaces. In contrast, SWDarkPeak exhibits weaker and more diffuse CO emission, with the SOFIA positions sampling comparatively faint molecular gas.

The spectral data reduction was performed following standard procedures using the CLASS package within the \textsc{Gildas} software.

\subsection{APEX \ci, $^{12}$CO(2-1), $^{13}$CO(2-1), $^{12}$CO(3-2), and $^{13}$CO(3-2)}

The observations of the CO rotational (J$=$2-1) transition at 230.538 GHz (rest frame) are presented in \citep{Saldano2023} and include N22, the SWbar regions, and the SWDarkPeak region in the SMC Bar, and N83 in the SMC Wing, with fields of view of 24.8\arcmin$\times$19.7\arcmin, 20.9\arcmin $\times$24.6\arcmin, 5.5\arcmin$\times$5.8\arcmin, and 4.1\arcmin$\times$13.8\arcmin, respectively. The rms noise of the data ranges between $\sim$0.1 and 0.7 K, the spatial resolution is $\sim$9 pc ($\sim30\arcsec$), and the smoothed velocity resolution is $\Delta V=0.25$ km s$^{-1}$. CO(2-1) spectra were extracted at the position of the SOFIA regions.

Observations of the neutral atomic carbon fine structure transition \ci\ ($^3$P$_1-^3$P$_0$) at 492.1607 GHz towards 13 pointings were done with the APEX telescope using the FLASH+ facility receiver as part of projects C-98.F-9706A-2016 and M-097.F-0025-2016 (PI Rúbio). These included the SMC regions; N22-R1, N22-R2, N22-R3, SWbarN-R4, SWbarN-R5, SWbarN-R6, N83-R1, N83-R2, N83-R3, N83-R4, during two observing runs in 2016. FLASH+ is a dual-frequency heterodyne receiver, operating simultaneously in the 345 and 460 GHz atmospheric windows, providing 4 GHz bandwidth in each sideband \citep{Klein2014}. The \ci (1-0) 492 GHz line was placed in the upper sideband of the high frequency channel, and the low frequency channel was tuned to 492.160 GHz to simultaneously cover the $^{13}$CO(3-2) and ${12}$CO(3-2) lines at 330.588 GHz and 345.796 GHz in the lower and upper sidebands, respectively. We used the XFFTS backends, providing a 38 kHz (0.033 km s$^{-1}$) spectral resolution for both the $J=3-2$ lines and 76 kHz (0.050 km s$^{-1}$) for the \ci (1-0) line. The data were smoothed to a velocity resolution of $\sim$1-2 km s$^{-1}$ and the rms is 0.035 K. 

More recent neutral carbon observations  were performed in the SWDarkPeak with APEX as part of project C-0116.F-9708C-2025 (PI Rodriguez-Ambler) using the nFLASH460 receiver.
The \ci (1-0) spectrum line at 492 GHz was obtained in October 2025 with an rms = 0.1 K  and a velocity resolution of $\Delta V = 0.4 $ km s$^{-1}$. We include this new observation in the discussion of the DarkPeak in Section \ref{darkpeak}.

$^{12}$CO(3-2) and $^{13}$CO(3-2) were simultaneously observed in some of the regions as shown in Table \ref{regions table}. 

   \begin{figure*}[!t]
   \centering
   \includegraphics[width=\hsize]{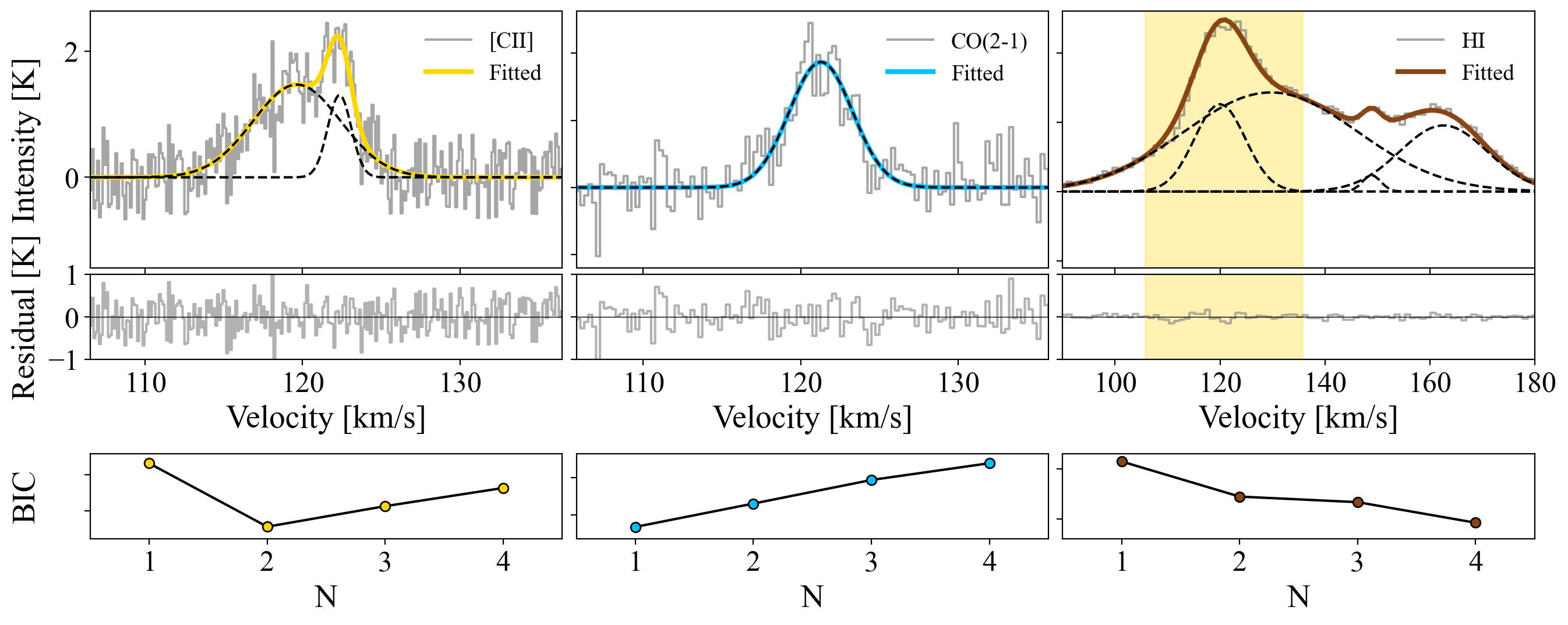}
      \caption{Multi-Gaussian fits to the \cii\ 158~$\mu$m (left), CO(2--1) (middle), and \hi\ (right) emission lines (gray) of N22-R1. Individual Gaussian components are shown as black dashed lines, and the fit residuals are shown in the middle panels. The bottom panels show the Bayesian information criterion, BIC, as a function of the number of Gaussian components, N. For the \hi\ spectrum, we consider a broader velocity interval, 80--200~km~s$^{-1}$, within which the BIC analysis identifies four distinct kinematic components. The component centered at $\sim$125~km~s$^{-1}$ coincides with the velocity range of the \cii\ and CO emission shown in the spectra (yellow shaded area).}

         \label{BIC}
   \end{figure*}

In this work, we only use the CO(2-1) observations, and a detailed analysis of the \ci\ data, as well as the CO(3-2) and isotopes, will be presented in a forthcoming paper.

\subsection{ASKAP+Parkes \hi}

 {We make use of the neutral atomic hydrogen \hi\ 21 cm data cube of the Small Magellanic Cloud obtained as part of the Galactic Australian Square Kilometre Array Pathfinder \hi\ (GASKAP-\hi) pilot survey \citep{Pingel2022}.The observations cover approximately 25 deg$^2$ of the SMC, with an angular resolution of $30\arcsec$ (corresponding to $\sim10$ pc at the distance of the SMC) and a spectral resolution of
$0.98~\mathrm{km~s^{-1}}$ in the Local Standard of Rest Kinematic
(LSRK) frame. The \hi\ data cube has an rms noise level of
$\sim1.1$ K per channel \citep{Pingel2022}. \hi\ spectra were
extracted at the positions of the SOFIA \cii\ 158 $\mu$m
pointings by sampling the 21 cm data cube at the corresponding sky
coordinates.}

\subsection{Sample selection} \label{sample selection}

We have 17 SOFIA pointings in total, of which 13 have simultaneous \cii\, CO(2–1) and \hi\ observations. Of these, 10 satisfy our signal-to-noise ratio (S/N) and velocity-shift criteria (Section \ref{line widths}) and are used in the decomposition analysis (Section \ref{decomposition}); the remaining three: SWDarkPeak, N83-R2 and N83-R4, are discussed individually in Section \ref{special cases} and \ref{darkpeak}.

\section{Infrared luminosity and star formation rate of the SMC} \label{LTIR and SFR}

Interpreting the \cii\ emission in the SMC requires a careful characterization of both the dust-heated infrared output and the recent star formation activity. We therefore adopt infrared luminosities derived from full dust SED fitting and SFR estimates based on recombination-line emission, providing a consistent description of the physical conditions across the galaxy.

\subsection{Infrared luminosity} \label{LTIR}

{The total infrared (TIR) luminosity, $L_{\rm TIR}$, was estimated following \citet{Draine2007},}

\begin{equation}
L_{\mathrm{TIR}} = P_0M_{\mathrm{dust}}\langle U \rangle,
\end{equation}

 {where $M_{\mathrm{dust}}$ is the dust mass, $\langle U \rangle$ is the dust-mass-weighted mean starlight intensity, and $P_0$ is the power absorbed per unit dust mass for a radiation field of intensity $U=1$. The dust mass and radiation field maps were taken from \citet{Chastenet2019}, who derived these quantities by fitting the Draine \& Li (2007, hereafter DL07) dust models using the Milky Way ($R_V=3.1$) grain mixture, following the methodology of \citet{Aniano2012}. Although the SMC exhibits dust properties that differ from those of the Milky Way, the MW grain model has been shown to provide satisfactory fits in low-metallicity environments while allowing the PAH mass fraction ($q_{\mathrm{PAH}}$) to remain a free parameter \citep{Draine2007,Sandstrom2010,Chastenet2019}. Consequently, the normalization $P_0$ must be consistent with the grain model adopted to derive the dust parameters rather than with the intrinsic dust-to-gas ratio of the SMC. We therefore adopt $P_0 = 120L_\odot M_\odot^{-1}$, corresponding to the MW $R_V=3.1$ grain model and consistent with the empirical calibration of \citet{Magdis2012}.}

\subsection{Special case of region SWbarN}

The \citet{Chastenet2025} dust parameter maps  for the SMC contain some empty pixels (NaN) in the SWbarN region. Therefore, for SWbarN-R4, SWbarN-R5, and SWbarN-R6, we adopted an alternative approach to estimate $\Sigma_{\mathrm{SFR}}$. Instead of correcting the H$\alpha$ luminosity with $L_{\mathrm{TIR}}$ derived from the DL07 mixture, we used the 160~$\mu$m emission map (12") from the \textit{Herschel} HERITAGE survey, which covers the SMC over $5^\circ \times 5^\circ$ (SPIRE) and $4^\circ \times 3^\circ$ (PACS) fields \citep{250microns}, to obtain $L_{\mathrm{TIR}}$ and calculate the obscured star formation rate.

To estimate the TIR luminosity from the \textit{Herschel} emission in the 160 $\mu$m band, we used the expression derived by \citet{Galametz2013}:

\begin{eqnarray}
    log\:L_{TIR}=1.024\:log\:\nu L_\nu(160\mu m)+0.176, \label{galametz}
\end{eqnarray}
with $a_{160}=1.024$ and $b_{160}=0.176$ the calibration coefficients, and $\nu L_{\nu}(160\mu m)$ the flux in the 160 $\mu m$ band in $L_\odot$.

 {As a consistency check, we also estimated $L_{\mathrm{TIR}}$ from the \textit{Herschel} 160~$\mu$m emission using the empirical calibration of \citet{Galametz2013} (Equation \ref{galametz}) for all regions where the DL07 dust-model parameters are available. The two methods yield consistent infrared luminosities, with a median ratio $  {L_{\mathrm{TIR}}^{\mathrm{DL07}}/L_{\mathrm{TIR}}^{160\mu m} \approx 1.3}$ and a scatter of 0.22 dex.}

\subsection{Star formation rate} \label{SFR}

We measure the local star formation rate (SFR) in the SMC following the calibration by \citet{KennicuttEvans2012}: 

\begin{equation}
\log {\rm SFR}~[M_\odot~\mathrm{yr}^{-1}] = \log L_{\rm H\alpha} - \log C_{\rm H\alpha},
\label{SFR}
\end{equation}

\noindent where the SFR calibration coefficient is $\log C_{\rm H\alpha} = 41.26$ \citep{Hao2011} assuming a \citet{Kroupa2001} IMF. The H$\alpha$ line emission map for the SMC comes from the UM/CTIO Magellanic Cloud Emission-line Survey that covers the central $3.5^\circ \times 4.5^\circ$ of the SMC at 3\arcsec–4\arcsec\ resolution with a photometric accuracy of $\sim$5\% \citep{Halpha}.

To account for the obscured star formation traced by the dust continuum emission, we correct the H\(\alpha \) luminosity by combining it with the total infrared luminosity, following \citet{Kennicutt2009}:

\begin{equation}L_{\mathrm{H}\alpha,\mathrm{corr}} = L_{\mathrm{H}\alpha,{\mathrm{obs}}} + 0.0024 L_{\mathrm{TIR}}.\label{eq:halpha_corr}\end{equation}

The calibration coefficient adopted in Eq.~\ref{eq:halpha_corr} was derived from samples dominated by galaxies with higher metallicities than the SMC \citep{Kennicutt2009}. Consequently, the TIR contribution may underestimate the obscured star formation component in the SMC.

\begin{figure}[!t]
\centering
\includegraphics[width=\hsize]{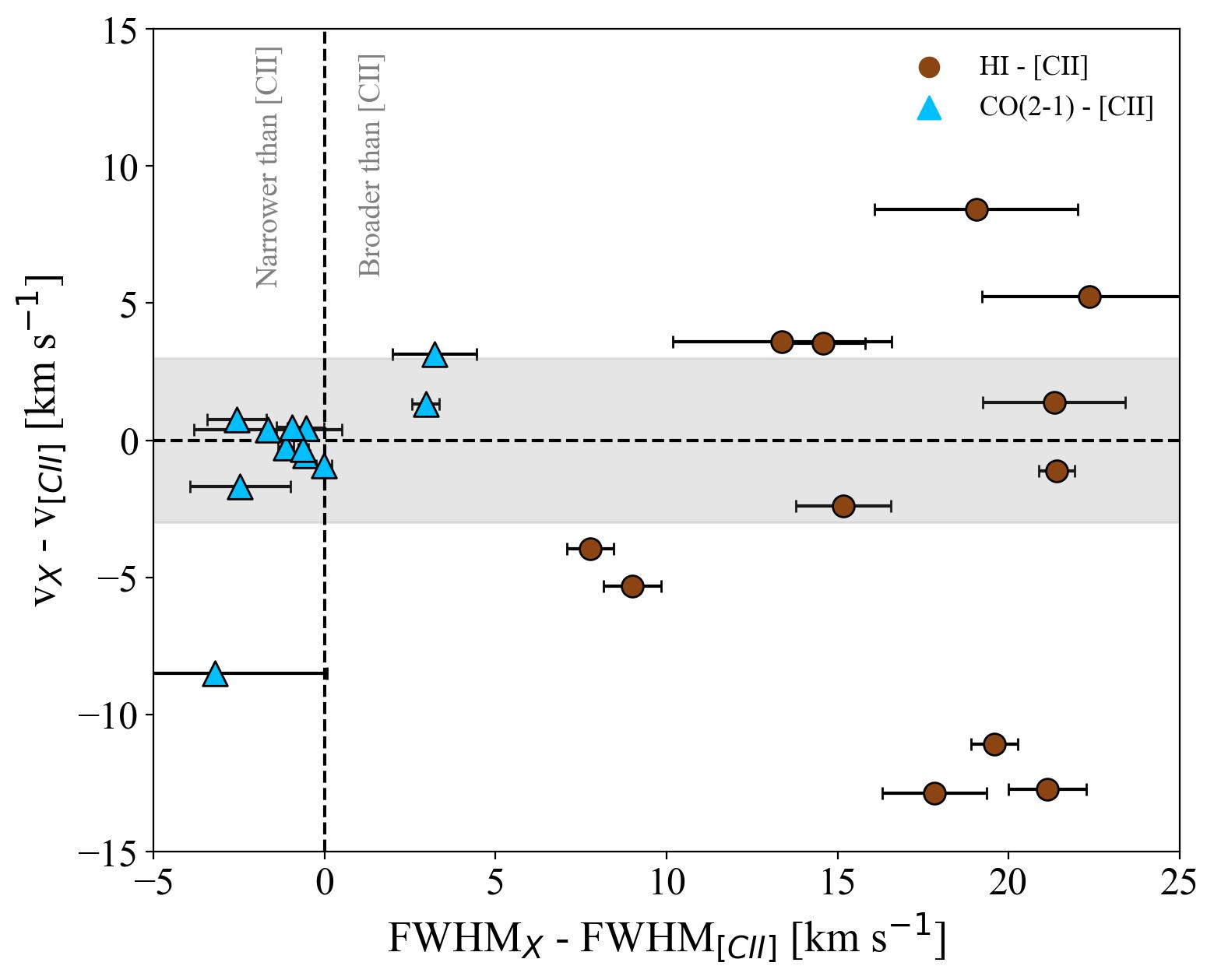}
\caption{Line width comparison among the \cii\ 158~$\mu$m, CO(2–1) (circles), and \hi\ (triangles) emission line profiles. Points lying on the left side of x-axis correspond to regions where the tracers' line width is narrower than \cii. Meanwhile, points on the right side, indicate regions where the CO and \hi\ are broader than \cii. The gray shaded area represents the velocity-shift tolerance range described in Section \ref{line widths}.}
\label{FWHM}
\end{figure}

   \begin{figure}[!t]
   \centering
   \includegraphics[width=\hsize]{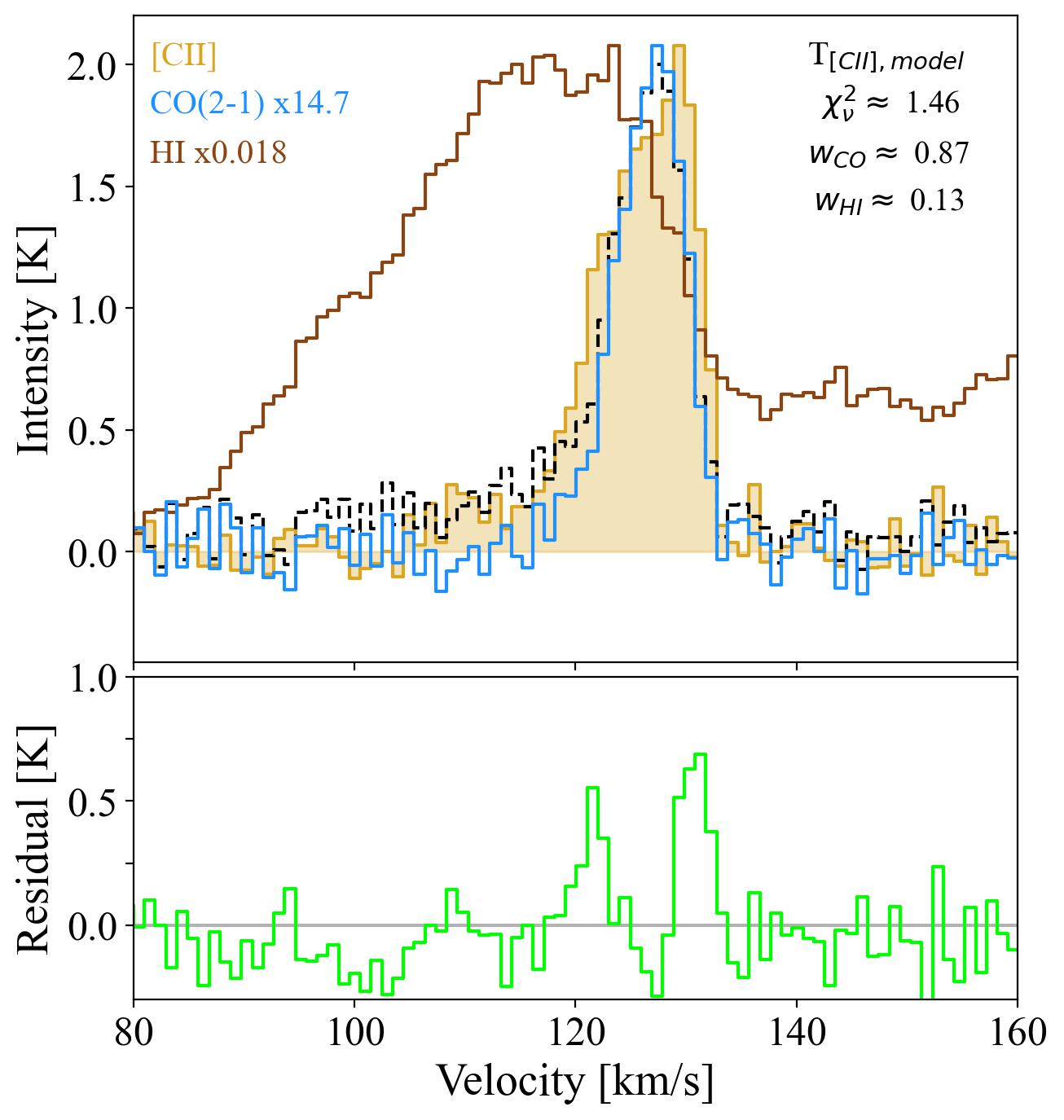}
    \caption{Rebinned spectra of \cii\ 158~$\mu$m (yellow) and CO(2--1) (light blue) for SWbarS-R5, matched to the velocity resolution of the \hi\ spectrum (brown). The \hi\ and CO spectra are normalized to the peak intensity of the \cii\ spectrum. The black dashed line shows the best-fit \cii\ 158~$\mu$m emission obtained from a linear combination of the CO and \hi\ spectra, with coefficients $w_{\rm CO}\approx0.87$ and $w_{\rm HI}\approx0.13$, and a reduced $\chi^2_\nu\approx  {1.46}$, following the methodology described in Sect.~\ref{decomposition}.}

         \label{[CII] model}
   \end{figure}

\section{Results} \label{results}

\subsection{Comparing velocities and line widths across tracers} \label{line widths}

Spectral line widths were measured by fitting Gaussian profiles to the observed emission lines and deriving the corresponding FWHM or velocity dispersion \citep[e.g.,][]{Muller2010, Pineda2017}. This provides a homogeneous framework for comparing the kinematic properties of different gas tracers in the low-metallicity interstellar medium of the SMC \citep[e.g.,][]{Rubio2015}.

\begin{figure*}[!t]
   \centering
    \begin{subfigure}{0.47\textwidth}
        \centering
        \includegraphics[width=\linewidth]{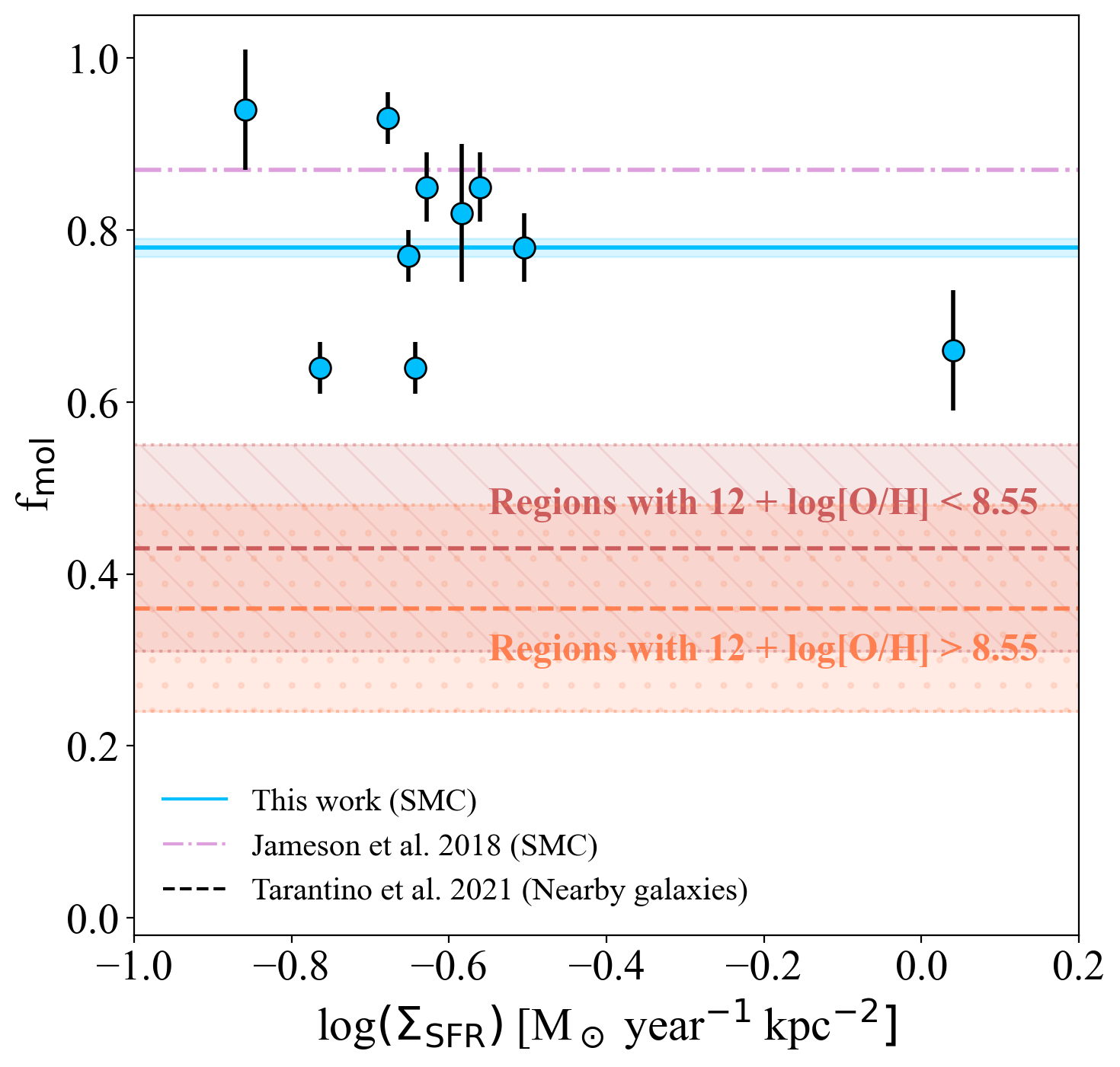}
        \caption{}
        \label{f_mol}
    \end{subfigure}
    \hfill
    \begin{subfigure}{0.47\textwidth}
        \centering
        \includegraphics[width=\linewidth]{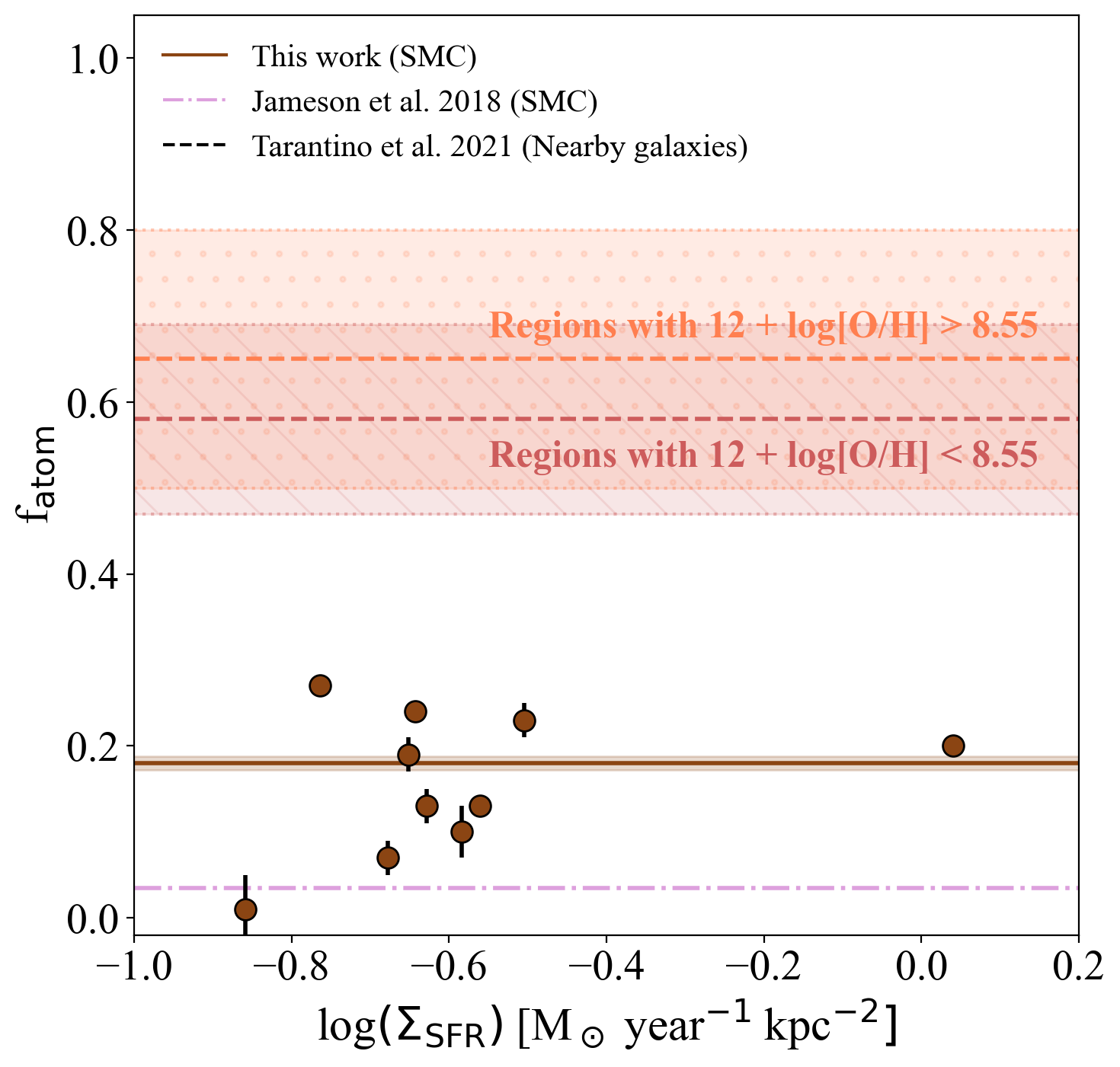}
        \caption{}
        \label{f_atom}
    \end{subfigure}
    \hfill

\caption{Estimated \cii\ 158~$\mu$m emission fractions associated with (a) CO-bright H$_2$ molecular gas and (b) \hi\ atomic gas, based on the \cii\ linear decomposition. The average \cii\ emission fractions across the SMC regions analyzed in this study are indicated by the light-blue and brown solid lines, respectively, with shaded areas representing the corresponding weighted uncertainties. Both panels also show the \cii\ 158~$\mu$m emission fractions estimated by \cite{Tarantino2021} for nearby galaxies (clay and orange dashed lines) and by \citet{Jameson2018} for the SMC (pink dash-dotted line), with their respective uncertainty bands, associated with the different gas phases.}

    \label{fractions image}
   \end{figure*}

Owing to the high velocity resolution of the available data (CO(2--1): $\sim0.25$ km s$^{-1}$, \cii: $\sim0.1$ km s$^{-1}$, and \hi: $\sim0.9$ km s$^{-1}$), we are able to resolve narrow and complex features in the line profiles, particularly in the \cii\ 158~$\mu$m emission. The top left panel of Figure \ref{SMC regions} shows a comparison of the line profiles observed toward the N22-R1 region. The top panel presents the \cii\ spectrum, which exhibits a complex velocity structure with asymmetric wings, indicating the presence of multiple kinematic components that cannot be adequately described by a single Gaussian function, as commonly found in low-metallicity star-forming regions \citep[e.g.,][]{HerreraCamus2015, Pineda2017}. In contrast, the CO and $^{13}$CO transitions display comparatively narrower and more symmetric profiles, tracing the denser molecular gas \citep[e.g.,][]{Rubio2015}. The \hi\ emission is significantly broader reflecting the widespread character and large velocity dispersion of the atomic gas, particularly its warm component \citep[e.g.,][]{Stanimirovic2004, Pingel2022}. These differences highlight the distinct kinematic regimes probed by each tracer and motivate a careful and homogeneous treatment of the linewidth measurements across gas phases.

To identify possible asymmetries, we perform multi-Gaussian fits and apply the Bayesian information criterion (BIC) to assess whether additional components significantly improve the fit. This approach allows us to identify spectra with deviations from a single Gaussian profile, while still adopting a FWHM as a uniform metric across the dataset. Figure~\ref{BIC} shows an example of this procedure for the region N22-R1. The middle panels present the resulting fits for the \cii\ (gold), CO(2–1) (light blue), and \hi\ (brown) native spectra (gray), together with the individual Gaussian components (black dashed lines). The middle panels show the corresponding residuals, meanwhile the bottom panels display the BIC as a function of the number of Gaussian components, where the minimum value indicates the preferred model complexity.

In this example, the BIC favors a multi-component description for the \cii\ profile, while the CO(2–1) spectrum is well reproduced by a single Gaussian component. The \hi\ line, in contrast, exhibits a broader and more structured profile requiring multiple components. N22-R1 therefore illustrates a behavior commonly observed across the SMC regions, where the \cii\ 158 $\mu$m emission line displays a more complex kinematic structure than the CO(2–1) emission.

To determine which component of the multi-Gaussian \cii\ fit should be compared with the CO and \hi\ components, we define a velocity matching criterion based on the relative peak positions. Specifically, we adopt a tolerance range for the velocity offset between the \cii\ peak and the CO and \hi\ peaks of $\Delta v \sim 3$~km~s$^{-1}$, corresponding to approximately 50\% of the average single-component linewidth. This threshold allows velocity differences comparable to the channel width of the \hi\ spectrum while remaining sufficiently restrictive to avoid associating clearly distinct kinematic components.

The results of the velocity and line width comparison are presented in Figure~\ref{FWHM}, where the gray shaded region indicates the adopted velocity-shift tolerance relative to the \cii\ centroid. We find that the \cii\ line widths systematically lie between those of CO(2-1) and \hi, with CO typically exhibiting narrower profiles and \hi\, broader ones. This trend supports a scenario in which \cii\ traces a combination of gas phases, arising from both the dense molecular component associated with CO emission and a more diffuse or extended component linked to atomic gas (\citealt{Okada2015}; \citealt{RequenaTorres2016}; \citealt{Okada2019}). The distribution of points in the FWHM comparison further reinforces this interpretation. Most CO measurements lie close to or below $\Delta \mathrm{FWHM} = 0$, indicating that CO traces the kinematically colder and more confined molecular cores. In contrast, \hi\ exhibits systematically larger line widths, consistent with a more turbulent and spatially extended medium. The intermediate position of \cii\ suggests that a significant fraction of its emission originates in molecular gas not traced by CO, commonly referred to as CO-dark H$_2$ (\citealt{Wolfire2010}; \citealt{Pineda2013}; \citealt{Bolatto2013}). 
\subsection{Decomposition analysis of the \cii\ transition} \label{decomposition}

To isolate the CO-faint molecular gas traced by the \cii\ line emission, we first need to remove the contributions arising from the ionized and atomic gas phases. In low-metallicity star-forming regions, the \cii\ emission associated with ionized gas is generally small compared to that originating in photodissociation regions (PDRs; \citealt{Kaufman1999}). Previous estimates of the ionized gas contribution to the total \cii\ emission are typically $\leq19$\% in the LMC \citep[e.g.,][]{Lebouteiller2012, Okada2015, Pineda2017} and $\leq5$\% in the SMC \citep[e.g.,][]{Pineda2017, RequenaTorres2016}. \cite{Jameson2018} presents {\em Herschel} PACS and SPIRE observations of these same regions in \cii\ and \nii\ emission and discusses the ionized gas contribution to \cii, concluding that it is at most 5\%. 
For these reasons we neglect the contribution of ionized gas to the \cii\ line emission of the SMC in this work.

To remove the atomic gas contribution to the \cii\ emission, we apply the method introduced by \citet{Tarantino2021} for nearby galaxies, in which the \cii\ line profile is modeled as a linear combination of the CO and \hi\ line profiles, representing the molecular and atomic contributions, respectively. For the decomposition, we use Rayleigh--Jeans brightness temperatures as a measure of flux for the \cii, CO($J$=2--1), and \hi\ spectra:

\begin{equation}
T_{\text{\cii},\mathrm{model}} = w_{\mathrm{CO}}T_{\mathrm{CO}} + w_{\text{\hi}}T_{\text{\hi}},
\label{linear decomposition}
\end{equation}

\noindent where $w_{\mathrm{CO}}$ and $w_{\text{\hi}}$ are the linear coefficients that best reproduce the observed \cii\ emission, and $T_{\text{\cii},\mathrm{model}}$ is the modeled \cii\ profile.  {For the \hi\ data, we consider the total multi-component spectrum at the velocity range where the \cii\ emission is located.}

The coefficients are obtained through $\chi^2$ minimization:

\begin{equation}
\chi ^2 = \sum_{n=1}^{N}\frac{\left(T_{\text{\cii}}-w_{\mathrm{CO}}T_{\mathrm{CO}}-w_{\text{\hi}}T_{\text{\hi}}\right)^2}
{\sigma_{\text{\cii}}^2+ w_{\mathrm{CO}}^2 \sigma_{\mathrm{CO}}^2+ w_{\text{\hi}}^2 \sigma_{\text{\hi}}^2},
\label{chi2}
\end{equation}

\noindent where $\sigma$ represents the rms noise of each rebinned spectrum, and the model is evaluated across $N$ spectral channels. Once the linear coefficients are obtained, we calculate the fractions of \cii\ emission associated with the molecular ($f_{\mathrm{mol}}$) and atomic ($f_{\mathrm{atomic}}$) phases:

\begin{equation}
f_{\mathrm{mol}} = \frac{w_{\mathrm{CO}}\int T_{\mathrm{CO}dv}}{\int T_{\text{\cii}}dv},
\qquad
f_{\mathrm{atomic}} = \frac{w_{\text{\hi}}\int T_{\text{\hi} dv}}{\int T_{\text{\cii}}dv}.
\label{fractions}
\end{equation}
 {The integration limits are defined from the Gaussian fit to the observed \cii\ line. Specifically, we integrate over the velocity interval encompassing the fitted \cii\ emission profile and apply the same limits to the CO($J$=2--1) and \hi\ spectra.}

Since the spectral lines were observed with different instruments, we rebinned the line profiles to a common velocity resolution $\Delta V \approx  {0.9}$ km s$^{-1}$ to perform the analysis described above. For each \hi\ velocity channel, we compute the mean \cii\ and CO emission over the corresponding velocity range. This approach enables a direct, channel-by-channel comparison for estimating the linear coefficients. Regarding the spatial resolution, \hi\ resolution is significantly poorer than that of the \cii\ and CO (about a factor of $\sim$ {3}). We decided not to convolve the \cii\ and CO maps to the \hi\ spatial resolution in order to preserve the level of detail of the analysis. We do not expect this to significantly impact the results, as the \hi\ spatial distribution is expected to be smoother than that of \cii\ and CO (See Figure \ref{BIC}).

An example is shown in Figure \ref{[CII] model}, where the solid lines represent the rebinned \cii\ (gold), CO (light blue) and \hi\ (brown), meanwhile the dashed black line corresponds to the modeled \cii\ emission. The \cii\ line profile lies between the \hi\ and CO profiles, a behavior present in 9 out of the 10 analyzed regions. In addition, excess \cii\ 158 $\mu$m emission at velocities that cannot be associated with CO is present in 8/10 of the regions (See Appendix \ref{[CII] models}). 

The \cii\ decomposition indicates that the gas associated with the CO-emitting molecular phase contributes, on average, $(  {78} \pm 1)\%$ of the total \cii\ emission, dominating over the atomic contribution. This result is illustrated in Figure \ref{fractions image}, which also compares our measurements with those of \citet{Jameson2018} for the SMC and \citet{Tarantino2021} for two nearby galaxies. Our results are in very good agreement with the SMC measurements of \citet{Jameson2018}. In contrast, \citet{Tarantino2021} find a more balanced molecular and atomic contribution to the \cii\ emission. We attribute this difference primarily to velocity and spatial resolution, and the metallicity of the studied regions. The analysis by \citet{Tarantino2021} has $\Delta V>5$ km s$^{-1}$ on $\sim 500$ pc scales in M101 and NGC6946, while our work is on scales of $\sim 4$ pc and a velocity resolution $\Delta V= 0.5$ km s$^{-1}$. Also, their regions are in environments with higher metallicities ($12 + log[\rm O/H]=8.31$–$8.62$ versus $12 + log[\rm O/H]_{SMC}\sim8.0$ for the SMC). These environmental conditions likely lead to a more evenly distributed molecular and atomic contribution to the \cii\ 158 $\mu$m emission.

This result provides evidence for CO-faint H$_2$ gas traced by \cii\ 158 $\mu$m emission in low-metallicity environments \citep{RequenaTorres2016, Langer2014, Pineda2014, Pineda2017}. In most regions, the residual between the modeled \cii\ emission profiles, based on the linear decomposition, and the observed spectra lies above the spectral noise level (see the bottom panel of Fig.~\ref{[CII] model}). Furthermore, the positive residuals are generally concentrated near the velocity range where CO emission peaks, suggesting that this excess \cii\ emission is kinematically associated with the molecular component rather than with unrelated atomic gas. This behavior is consistent with a scenario in which \cii\ traces molecular gas that shares the same dynamical structure as the CO-emitting material but remains weak or undetected in CO due to photodissociation effects in low-metallicity environments, where reduced dust shielding allows UV photons to penetrate deeper into molecular clouds, enlarging the C$^+$/H$_2$ layer surrounding dense CO cores \citep{Wolfire2010, Langer2014, Pineda2017}. Therefore, the residual emission provides additional evidence for an extended reservoir of CO-faint molecular gas surrounding the denser CO-bright regions.

Regarding the rebinning to a common spectral resolution between \cii, CO, and \hi, we do not expect this to introduce systematic effects unless the components were extraordinarily narrow, which we do not observe in the lines at native spectral resolution.

\begin{table}[]
\centering
\caption{Average H$_2$ molecular gas column density estimates, based on CO, \cii\ and CO+\cii, respectively, for the SMC regions.}
\label{column densities}
\resizebox{\columnwidth}{!}{%
\begin{tabular}{@{}lllll@{}}
\toprule\toprule
SMC region & $\overline{N}(\rm H_2)_{\rm CO}$ & $\overline{N}(\rm H_2)_{\text{\cii}}$  & $\overline{N}(\rm H_2)_{\rm total}$  \\ 
 & $[10^{21}\text{cm}^{-2}]$ & $[10^{21}\text{cm}^{-2}]$ &  $[10^{21}\text{cm}^{-2}]$\\
\midrule
N22 &$  {1.45 \pm 0.02}$  & $  {6.47\pm 0.01}$& $  {7.92 \pm 0.03}$   \\
SWbarN & $  {3.21 \pm 0.03}$ &  $  {11.6 \pm 0.02}$  & $  {14.8 \pm 0.03}$ \\
N83 & $  {1.91 \pm 0.07}$ & $  {4.60 \pm 0.02}$ &  $  {6.50 \pm 0.07}$  \\ 
SWbarS & $  {1.96 \pm0.02}$ &  $  {6.32 \pm 0.03}$  & $  {8.29 \pm 0.02}$ \\
\hline 
\end{tabular}
}
\end{table}

\begin{figure}
   \centering
   \includegraphics[width=\hsize]{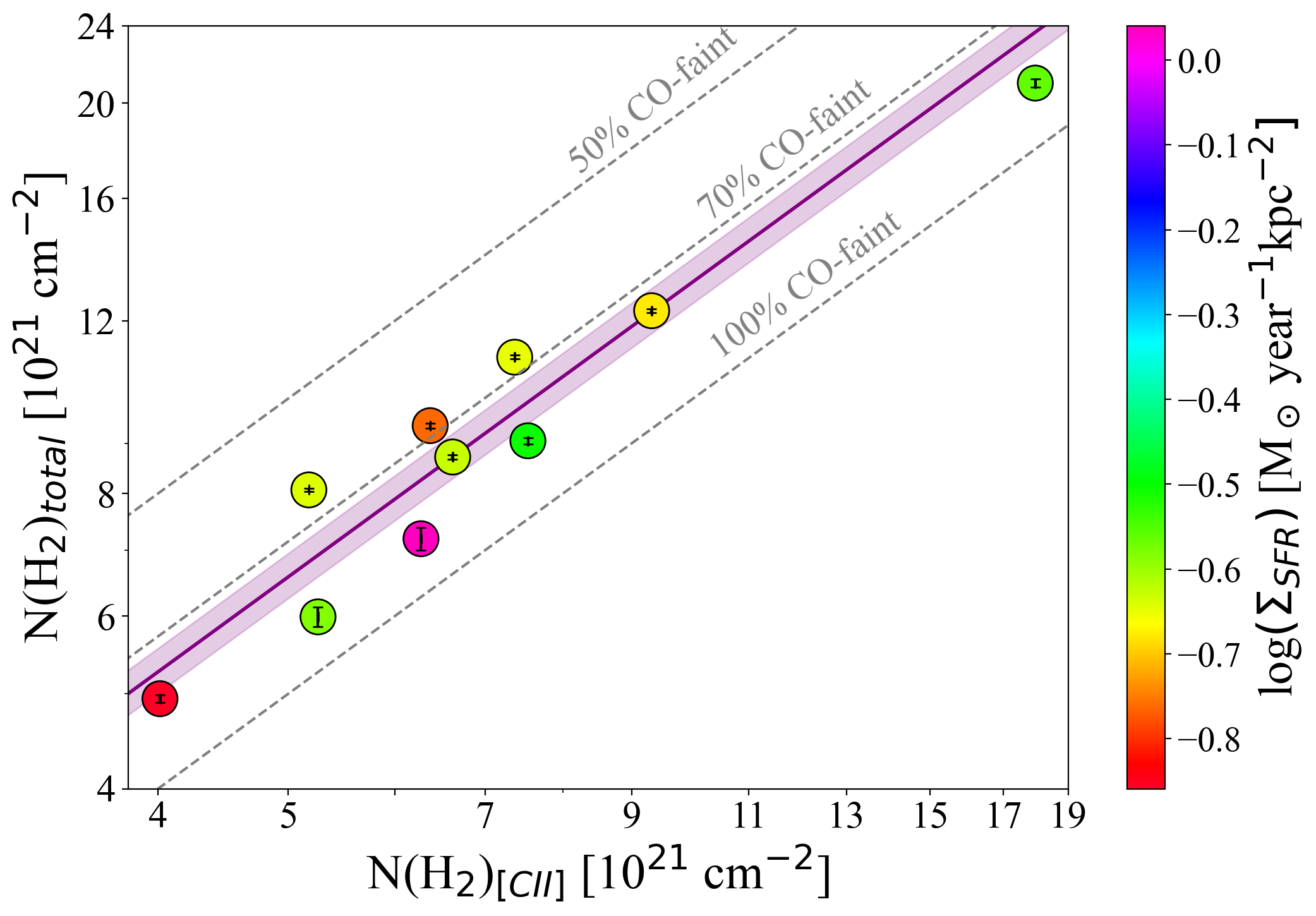}
     \caption{Total molecular hydrogen column density, $N(\mathrm{H}_2)_{\mathrm{total}}$, traced by \cii\ and CO across the analyzed SMC regions as a function of the molecular hydrogen column density traced by \cii, $N(\mathrm{H}_2)_{\text{\cii}}$. The error bars are listed in Table~\ref{column densities}. On average, \cii\ accounts for $\sim(77\pm4)\%$ of the total $N(\mathrm{H}_2)$ (purple line and shaded region), indicating that a substantial fraction of the molecular gas is in the CO-dark phase. This highlights the role of \cii\ as a tracer of molecular gas in low-metallicity environments such as the SMC. The color coding indicates the star formation rate surface density, $\log(\Sigma_{\mathrm{SFR}})$, in units of $M_\odot\,\mathrm{yr}^{-1}\,\mathrm{kpc}^{-2}$.}
         \label{CO dark gas}
   \end{figure}

\section{The CO-faint gas in the SMC: Tracing H$_2$ with CO and \cii} \label{CO faint gas}

In the previous section, we have shown that kinematically most of the \cii\ emission is associated with CO-bright molecular gas, while very little ($\lesssim20\%$) appears related to the \hi\ in emission. Moreover, the residuals have velocities that are strongly clustered around the CO emission, suggesting that they are either due to CO-faint molecular gas or possibly a cold phase of \hi\ that is enveloping the CO-emitting gas. 

Thus, due to the photodissociation of CO in the outer layers of an \hi/PDR/molecular cloud complex, it is necessary to consider the contributions of both CO and \cii\ emission to the total estimate of the H$_2$ column density:
\begin{equation}
N(\rm H_2)_{\rm total}=N(\rm H_2)_{\text{ \cii}}+N(\rm H_2)_{\rm CO},
\end{equation}
where $N(\rm H_2)_{\text{\cii}}$ and $N(\rm H_2)_{\rm CO}$ represent the molecular gas at low and high extinction, respectively. In this section, we derive $N(\rm H_2)$ independently from \cii\ and CO and quantify the fraction of molecular gas that is not traced by CO line emission in the low-metallicity environment of the SMC.

\subsection{Converting $I_{\text{\cii}}$ to $N(\mathrm{H_2})$} \label{I_CII to N(H2)}

Studies of \cii\ 158~$\mu$m emission in low-metallicity star-forming regions suggest that the \cii-emitting envelopes surrounding molecular clouds are generally optically thin for gas temperatures above $\sim11$ K, allowing the observed emission to be directly related to the column density of C$^+$ through collisional excitation models (LMC: \citealt{Israel1996}; SMC: \citealt{IsraelMaloney2011}). Following the formalism presented by \citet{Jameson2018}, the H$_2$ column density associated with the molecular component traced by \cii\ can be expressed as:

\begin{equation}
N(\mathrm{H_2})_{\text{ \cii}}=
\frac{4.35\times10^{23}}{(\rm C/H)_{\mathrm{SMC}}}
\left(
\frac{
1+2e^{-91.2/T}+n_{\rm crit}(\mathrm{H_2})/n
}{
2e^{-91.2/T}
}
\right)
I_{\text{\cii},\text{mol}},
\label{NH2_CII}
\end{equation}

where $I_{\text{\cii},\mathrm{mol}}$ is the integrated \cii\ intensity in unit of K km s$^{-1}$, associated with the molecular component ($f_{\rm mol}$), $n$ is the density of the collisional partner, $T$ is the gas temperature, and $n_{\rm crit}$ is the critical density for collisions with H$_2$ \citep{Crawford1985}. The carbon abundance adopted for the SMC is $(\rm C/H)_{\rm SMC}=2.8\times10^{-5}$, obtained by scaling the Galactic abundance according to the metallicity of the SMC \citep{Jameson2018}. 

The critical density for collisions with H$_2$ is:
\begin{equation}
n_{\rm crit}(\rm H_2)=\frac{A_{ul}}{R_{ul}}
\end{equation}
, where the collisional de-excitation rate coefficient for \cii-H$_2$ interactions, including hydrogen spin effects in LTE, is
$R_{ul} = (4.55 + 1.6e^{-100/T}) \times 10^{-10}$~cm$^{3}$~s$^{-1}$ \citep{WiesenfeldGoldsmith2014},
and $A_{ul} = 2.1\times10^{-6}$~s$^{-1}$ \citep{CarilliWalter2013}. As an example, the critical density at T$=$E$_{ul}/k\approx91.2$~K is therefore $n_{\rm crit}\approx4500$~cm$^{-3}$.

To convert the \cii\ emission into a molecular gas column density, we first use Eq.~\ref{NH2_CII} to calculate N(H$_2$)$_{\text{\cii}}$ for each SMC region from the corresponding integrated molecular \cii\ intensity, ($I_{\text{\cii,mol}}$), obtained from the decomposition analysis. The resulting column densities are then averaged to derive a mean value for the SMC. This conversion requires adopting representative gas conditions (density and temperature), which determine the \cii\ excitation.  {\citet{Jameson2018} constrained the properties of PDR gas using PDR Toolbox with updated SMC models, and find that the typical volume densities are $n\sim10^3-10^4$ cm$^{-3}$.} We assume an average density ($n = 4000$~cm$^{-3}$) and a temperature $T = 90$ K  {that resemble the excitation temperature and critical density of the transition \citep{WiesenfeldGoldsmith2014}}, following \citet{Jameson2018},  {and} consistent with warm and moderately dense PDRs. These values are based on the PDR models of \citet{Kaufman2006}, with updated gas and grain-surface chemistry from \citet{Wolfire2010} and \citet{Hollenbach2012}. At this temperature and density, the inferred H$_2$ column densities are approximately a factor of three larger than those obtained under the high-density, high-temperature limit corresponding to maximum \cii\ excitation. 
 {To evaluate the sensitivity of the derived column densities to the adopted density, we calculated N(H$2$)$_{\text{\cii}}$ assuming $n=3000$~cm$^{-3}$, while keeping $T=90$ K fixed. The resulting column densities are higher than those obtained for the assumed density, $n=4000$~cm$^{-3}$, by an average of $\sim13\%$. This indicates that the adopted density introduces a small systematic uncertainty in the derived molecular gas column densities.}

\subsection{Converting $I_{\rm CO}$ to $N(\rm H_2)$}

In the inner parts of molecular clouds, where shielding from the incident FUV radiation is sufficient to prevent CO from being photodissociated \citep{Wolfire2010, Lee2015}, the molecular gas column density can be estimated using the CO integrated intensity method,
\begin{equation}
    N(\rm H_2)=X_{\rm CO}I_{\rm CO},\label{X_CO} 
\end{equation}
where $N(\rm H_2)$ is the H$_2$ column density, $X_{\rm CO}$ is the CO-to-H$_2$ conversion factor in units of cm$^{-2}$ (K km s$^{-1}$)$^{-1}$, and $I_{\rm CO}$ is the rebinned CO integrated intensity in K km s$^{-1}$.

Although the low metallicity of the SMC is expected to increase the effective CO-to-H$2$ conversion factor, $X_{\rm CO}$, owing to the presence of extended CO-faint molecular envelopes \citep[e.g.,][]{Leroy2011}, simulations by \citet{Shetty2011} have shown that, in the densest and CO-bright regions of molecular clouds in the SMC, the Galactic conversion factor estimated by \citet{Bolatto2008}, $X_{\rm CO,MW} \approx 2\times10^{20}$ cm$^{-2}$ (K km s$^{-1}$)$^{-1}$, which is the canonical value for CO(1-0) assuming a CO line ratio $R_{21}=1$, can still provide a reasonable approximation. Furthermore, \citet{Jameson2018} estimated the conversion factor for CO-bright regions in the SMC and obtained median values of $X_{\rm CO} = (1.3$-$2.3)\times10^{20}$ cm$^{-2}$ (K km s$^{-1}$)$^{-1}$, consistent with a Milky Way value of $X_{\rm CO,MW}$. Given these simulation results, we adopt a Galactic CO-to-H$2$ conversion factor of $X_{\rm CO,MW}=2\times10^{20}$ cm$^{-2}$~(K km s$^{-1}$)$^{-1}$ \citep{Bolatto2013} to estimate the molecular gas content in regions of high  extinction where most of the carbon is in the form of CO. This choice is consistent with a scenario in which the main driver of the enhanced effective conversion factor in low-metallicity systems is the shrinking of the high-$A_V$ CO-emitting cores and the growth of an outer H$_2$ layer coextensive with C$^+$ \citep[e.g.,][]{Wolfire2010, Bolatto2013}.

\subsection{Comparing the \cii- and CO-based $N(H_2)$}

In Figure \ref{CO dark gas} we show that the molecular hydrogen traced by \cii, N(H$_2$)$_{\text{\cii}}$, accounts on average for roughly $77\%$ of the total molecular gas column density, N(H$_2$)$_\mathrm{total}$. The column density values derived from CO, \cii, and their combined contribution (CO+\cii) for each region are reported in Table \ref{column densities}. This corroborates that a substantial fraction of the molecular reservoir of the SMC is in the CO-faint phase, confirming results reported for the SMC using the same technique with {\em Herschel} PACS measurements \citep{Jameson2018}, as well other results using velocity-resolved \cii\ data for low-metallicity systems \citep[e.g.,][]{Pineda2014,Pineda2017}. The dominance of \cii–bright, CO-faint gas emphasizes that the H$_2$ content inferred from CO alone would underestimate the true molecular mass by almost a factor of $\sim$4 under low-metallicity conditions. 

Figure \ref{CO dark gas} also shows, however, that no significant correlation is found between N(H$_2$)${_{\text{\cii}}}$/N(H$_2$)$_{\rm total}$ and either the total molecular column density or the star formation rate surface density, $\Sigma_{\mathrm{SFR}}$. This lack of correlation suggests that the fraction of CO-faint gas is not mainly set by how dense or active in star formation a region is, but instead by how efficiently the gas and dust can shield molecules from the surrounding radiation field. Similar trends have been reported in other nearby dwarf galaxies (e.g. \citealt{Lebouteiller2012, Madden2020}), where \cii\ remains the main coolant of both quiet and active regions. These results indicate that in the SMC, the CO-dark molecular phase is widespread and relatively stable, and its presence does not depend strongly on the current level of star formation. 

The aforementioned is consistent with the physical picture proposed for low-metallicity dwarf galaxies by \citet{Cormier2019}, in which the reduced dust abundance produces a highly porous multiphase ISM, allowing far-ultraviolet radiation to penetrate deeper into molecular clouds. Under these conditions, the C$^+$/H$_2$ envelopes surrounding molecular clouds become spatially extended while the well-shielded CO cores shrink, increasing the fraction of molecular gas that remains invisible in CO emission. 

We can use these data to compute an $X_{\rm CO}$ factor applicable on $\sim4$~pc scales near CO-bright regions of the SMC. By solving the equation \ref{X_CO} for $X_{\rm CO}$, using $N(\rm H_2)$ as the total molecular gas column density, and $I_{\rm CO}$ as the weighted average of the integrated CO(2-1) intensity across the SMC regions, we estimate a CO(2-1) conversion factor $X_{\rm CO+CO_{dark}}=  {8.9}\times10^{20}$ cm$^{-2}$(K km s$^{-1}$)$^{-1}$. 
The Galactic conversion factor is typically calibrated for the CO(1-0) transition. Assuming a line ratio $R_{21}=\rm CO(2-1)/CO(1-0)\sim1$ for the SMC (\citealt{Rubio1996}), our result
is  {4.5} times larger than the canonical Galactic conversion factor ($X_{\mathrm{CO,MW}}\sim2\times10^{20}$ cm$^{-2}$(K km s$^{-1}$)$^{-1}$) \citep{Bolatto2008} reflecting the reduced CO abundance and enhanced CO–faint molecular gas expected in metal-poor environments.

Previous studies have shown that the CO-to-H$_2$ conversion factor in the SMC is significantly larger than in the Milky Way, although the inferred value depends strongly on the adopted tracer and spatial scale. Measurements based on virial masses of resolved molecular clouds typically yield $X_{\rm CO}\sim(4$--$7)\times10^{20}$ cm$^{-2}$ (K km s$^{-1}$)$^{-1}$ \citep{Bolatto2003}, while dust-based analyses that account for the extended CO-faint molecular reservoir find substantially larger values, reaching $\sim10^{21}$--$10^{22}$ cm$^{-2}$ (K km s$^{-1}$)$^{-1}$ across the SMC \citep{Rubio1996, Leroy2007, Leroy2011}. 
\citet{Leroy2011} used {\em Spitzer} observations to model the dust emission of the SMC and concluded that the $X_{\mathrm{CO}}$ factor on large scales is $\sim16$ times larger than in the Milky Way. \citet{Jameson2016} modeled {\em Herschel} dust mapping of the SMC to find that the molecular mass of the galaxy is $\sim17$ times larger than that derived from its CO luminosity and a Galactic conversion factor.
More recent studies combining \cii, \ci, and CO observations have further demonstrated that a substantial fraction of the molecular gas in low-metallicity systems is CO-dark, leading to effectively larger conversion factors when the total H$_2$ mass is considered (e.g., \citet{Pineda2017}, who found $X_{\mathrm{CO}}\sim10^{20}-10^{22}$ cm$^{-2}$ (K km s$^{-1}$)$^{-1}$ depending on the region). Our estimate of $X_{\rm CO}=  {8.9}\times10^{20}$ cm$^{-2}$ (K km s$^{-1}$)$^{-1}$ therefore lies within the range of values reported for individual molecular complexes in the SMC, while still reflecting the influence of a substantial CO–faint molecular component.

\subsection{N83-R2, N83-R4 and SWDarkPeak-R7} \label{special cases}
The tracing of the total molecular gas content in the SMC is assessed mainly through the \cii\ 158 $\mu$m emission of the selected regions, so we cannot carry this examination in N83-R4, since there is no detection of CO(2-1) in the position of the central pixel of the SOFIA/upGREAT array. To perform the linear decomposition of the \cii\ 158 $\mu$m emission line described in section 4.1 it is necessary to have spectra with at least $S/N \geq10$ \citep{Tarantino2021}. Since the kinematic intensity peak of CO emission line in N83-R2 is shifted from that of \cii\ by $\Delta v\approx 4$ km s$^{-1}$, out of our tolerance range, and the CO(2-1) spectrum of the SWDarkPeak has a $S/N\approx 4.6$, we did not include these regions in the CO-dark molecular gas estimation. 

\begin{table}[]
\centering
\caption{Carbon phase ratios for the 2 kinematic components along the LOS in the SWDarkPeak region.}
\label{carbon phase ratios}
\resizebox{\columnwidth}{!}{
\begin{tabular}{@{}lllll@{}}
\toprule\toprule
Component & $I_{\rm CO}/I_{\text{\cii},mol}$ & $I_{\text{\ci}}/I_{\text{\cii},mol}$  & $I_{\text{\ci}}/I_{\rm CO}$  \\ 
\midrule
SWDarkPeak-A & 0.12   &  -  &  -  \\
SWDarkPeak-B & 3.6 &  0.46  & 0.13 \\
\hline 
\end{tabular}
}
\end{table}

\section{The interesting case of SWDarkPeak} \label{darkpeak}

The SWDarkPeak region was found in the dust-based mapping of molecular gas in the SMC by \citet{Bolatto2011} and \citet{Jameson2016} as a likely molecular peak (based on its high dust-to-\hi\ ratio) with no known CO emission. Later CO mapping reported by \citet{Jameson2018} detected faint CO emission within the velocity range $\sim137$-160 km s$^{-1}$, and measured high \cii/CO ratios. The nature of the SWDarkPeak region makes it an ideal laboratory to study the transition between CO-dark and CO-bright molecular gas.  In contrast to other SMC regions, SWDarkPeak shows no associated bright H$\alpha$ emission, indicating a lack of significant massive star formation and suggesting that the observed \cii\ emission arises primarily from photodissociation regions associated with diffuse or weakly shielded molecular gas. 

\begin{figure}
    \centering
    \includegraphics[width=1\linewidth]{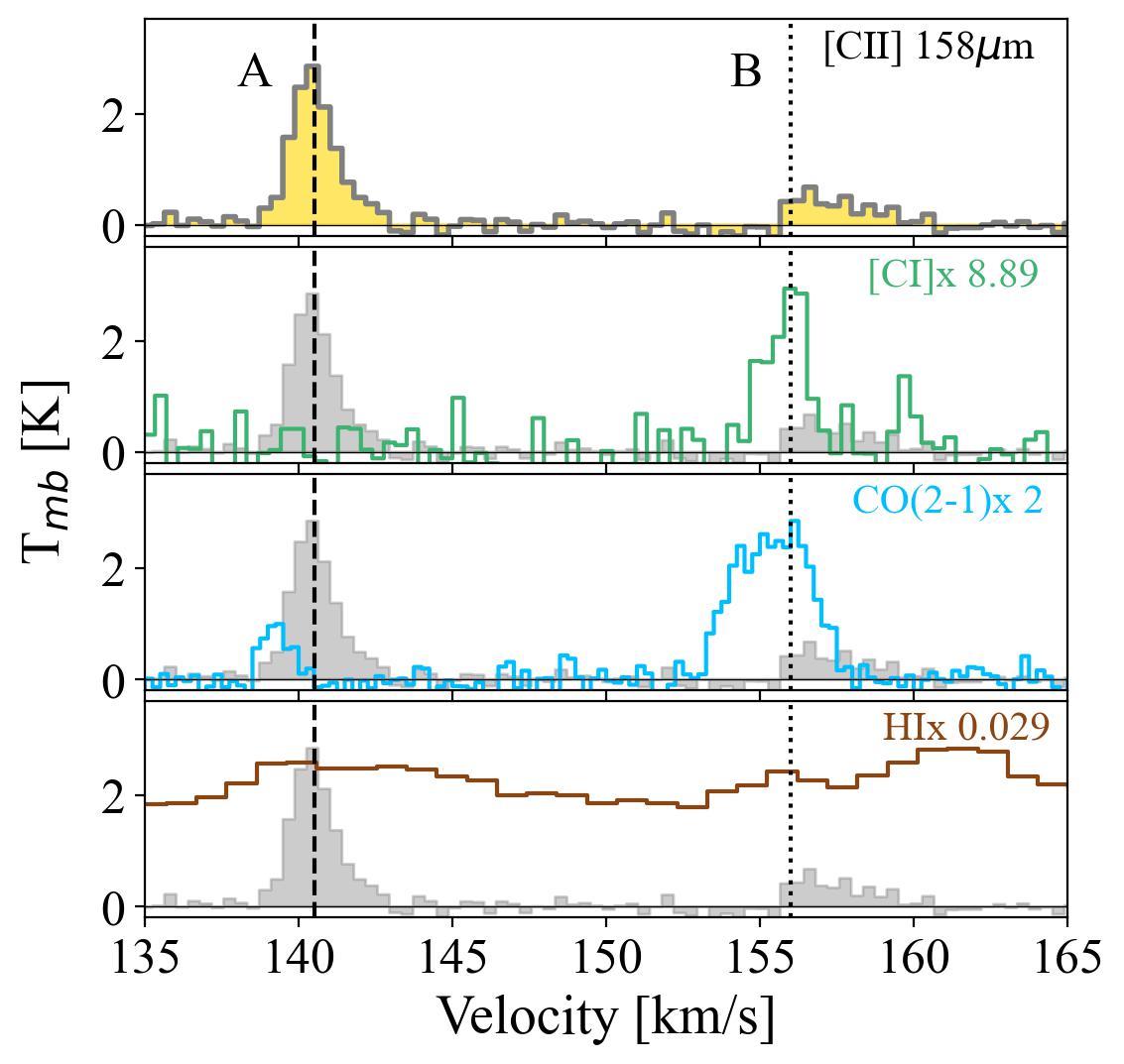}
    \caption{Spectral emission from the SWDarkPeak region in \cii\ (gold), \ci\ (green), CO(2--1) (light blue), and \hi\ (brown), normalized to the peak intensity of the \cii\ emission. The black dashed and dotted lines indicate the SWDarkPeak-A and SWDarkPeak-B components, respectively. The two components exhibit distinct molecular gas properties: SWDarkPeak-B is a CO-emitting molecular cloud with a $\sim40\%$ CO-dark molecular gas contribution, whereas SWDarkPeak-A has a $\sim90\%$ CO-dark molecular gas contribution.}
    \label{SWDP}
\end{figure}

We analyze this low-metallicity region to investigate the impact of PDR conditions on the ability of CO to trace molecular gas. In Fig.~\ref{SWDP}, we present the newly acquired \ci\ spectrum ($S/N \sim 7$), together with the existing \hi, CO, and \cii\ data. Thanks to the high spectral resolution of our observations, we identify two distinct kinematic components along the line of sight toward SWDarkPeak: SWDarkPeak-A (dashed black line) and SWDarkPeak-B (dotted black line). The strong \cii\ emission detected at $\sim140$ km s$^{-1}$, which lacks a corresponding \ci\ and strong CO counterpart, most likely arises from a CO-dark molecular phase in which hydrogen is predominantly molecular while carbon remains ionized due to reduced dust shielding in the low-metallicity environment (\citealt{Wolfire2010}; \citealt{Bolatto2013}; \citealt{Pineda2013}). In contrast, the CO(2-1) and \ci\ emission detected at $\sim156$ km s$^{-1}$ trace a more shielded molecular component in which carbon has largely transitioned to neutral and molecular (CO) form.

To probe the nature of these two kinematic components, we use our estimates of the molecular gas contribution to the \cii\ 158 $\mu$m emission ($\overline{f}_{\rm mol}\approx0.8$; see Figure \ref{fractions image}) to derive the corresponding molecular hydrogen column densities traced by \cii\ and CO. For the SWDarkPeak-A component ($v\sim140$ km s$^{-1}$), we obtain $N(\mathrm{H}_2)_{\text{\cii}} = (1.7\pm0.4)\times10^{21}$ [cm$^{-2}$] and $N(\mathrm{H}_2)_{\rm CO} = (1.0\pm 1.7) \times 10^{20}$ cm$^{-2}$, implying a CO-dark molecular gas fraction of $f_{\rm CO-dark} =0.9$. In contrast, for the SWDarkPeak-B component ($v\sim156$ km s$^{-1}$), we measure $N(\mathrm{H}_2)_{\text{\cii}} = (5.0\pm0.1)\times 10^{20}$ cm$^{-2}$ and $N(\mathrm{H}_2)_{\rm CO} = (8.3\pm0.6)\times 10^{20}$ cm$^{-2}$, corresponding to $f_{\rm CO-dark} =0.4$ . These results suggest that the SWDarkPeak line of sight intercepts two physically distinct molecular environments: a CO-dark molecular component traced primarily by \cii\ emission at $\sim140$ km s$^{-1}$ and a more shielded molecular cloud traced by CO and \ci\ at $\sim156$ km s$^{-1}$.

The simultaneous detection of \cii, \ci, and CO toward the SWDarkPeak-B component further allows us to probe the carbon phase balance within the molecular gas. We compare the integrated line intensities of the three tracers and derive the ratios $I_{\rm CO}/I_{\text{\cii},\rm mol}$,  $I_{\text{\ci}}/I_{\text{\cii},\rm mol}$ and $I_{\text{\ci}}/I_{\rm CO}$  for each velocity component. As shown in Table \ref{carbon phase ratios}, the SWDarkPeak-A component exhibits a very low $I_{\rm CO}/I_{\text{\cii},\rm mol}$ ratio together with an undetected \ci\ counterpart, indicating that the carbon reservoir is dominated by ionized carbon. In contrast, the SWDarkPeak-B component shows detected \ci\ and CO emission, suggesting that carbon has largely transitioned from C$^{+}$ to neutral and, mostly, molecular forms. This behavior follows the classical stratification expected in photodissociation regions, where the dominant carbon phase evolves from C$^{+}$ to C$^{0}$ and finally to CO with increasing shielding \citep{TielensHollenbach1985, HollenbachTielens1999}.

\section{Conclusions} \label{conclusions}

\begin{enumerate}

\item Using velocity-resolved SOFIA/GREAT and upGREAT \cii\ 158 $\mu$m, APEX, and  {ASKAP+Parkes} \hi\ 21 cm observations, we decomposed the \cii\ emission in several regions of the Small Magellanic Cloud into atomic and molecular components following the approach of \citet{Tarantino2021}. On average, $  ({78}\pm1)\%$ of the \cii\ luminosity arises from the molecular phase, while $(  {18}\pm1)\%$ is associated with atomic gas, confirming that \cii\ predominantly traces molecular gas in low-metallicity environments \citep[e.g.,][]{RequenaTorres2016,Langer2014,Pineda2014}.

\item The residuals of the \cii\ decomposition are mostly positive, indicating that the linear fitting approach tends to underestimate the \cii\ emission fractions, particularly from the molecular component. A small additional contribution ($\lesssim4\%$) may arise from ionized gas, as suggested by \citet{HerreraCamus2016} and \citet{Croxall2017}. Complementary \nii\ observations would be necessary to quantify this contribution in detail for the SMC.

\item Converting the decomposed \cii\ emission into column densities using the approach described in Section \ref{I_CII to N(H2)}, we find that the \cii\ traced H$_2$ accounts for roughly $(77\pm4)\%$ of the total molecular gas column density, with the remaining $\sim20\%$ traced by CO. This demonstrates that the bulk of the molecular gas in the SMC is CO-faint, consistent with results obtained for other low-metallicity systems \citep[e.g.,][]{Pineda2017, Jameson2018,Madden2020}. 

We find that the CO-to-H$_2$ conversion factor on $\sim4$~pc scales and in the proximity of CO emitting molecular complexes is $X_{\rm CO}\approx   {8.9}\times 10^{20}$ cm$^{-2}$ (K km s$^{-1}$)$^{-1}$ after accounting for the H$_2$ mixed with both the C$^+$ and the CO, about  {4.5} times larger than the canonical Galactic conversion factor.
\item We find no significant correlation between the CO-dark fraction, $N({\rm H_2})_{\text{\cii}}/N({\rm H_2})_{\mathrm{total}}$, and either the total molecular column density or the star formation rate surface density. This suggests that the abundance of CO-faint gas is not primarily driven by local density or star formation, but instead by the efficiency of dust and gas shielding against the FUV radiation field, which scales with metallicity (e.g., \citealt{Bolatto2013,Leroy2011,Okada2019}). Future work will explore the spatially resolved spectroscopic data from the \textit{Local Volume Mapper} (LVM) of SDSS-V \citet{Kollmeier2019} to investigate possible metallicity gradients across the SMC and their effect on the \cii/CO relation.

\end{enumerate}
Our study demonstrates that \cii\ is the dominant tracer of molecular gas in the SMC (77\%) and likely in other metal-poor systems. The CO-dark molecular phase represents an extended, stable, and significant component of the low-metallicity interstellar medium that is not primarily driven by stellar feedback, but rather by chemical processes. This provides an important clue for interpreting \cii\ emission in unresolved galaxies (e.g., \citealt{Cormier2019, Okada2019}).

\begin{acknowledgements}
R.H.-C. thanks the Max Planck Society for support under the Partner Group project "The Baryon Cycle in Galaxies" between the Max Planck for Extraterrestrial Physics and the Universidad de Concepción. R.H-C. and K.R-A. also gratefully acknowledge financial support from ANID - MILENIO - NCN2024\_112 and ANID BASAL FB210003
M.R. and K.R-A wishes to acknowledge partial support from ANID (CHILE) through Basal FB210003 and through FONDECYT grant No1190684. We also wish to thank Mariana Muñoz for reducing the CO(2-1) APEX  data and Jérémy Chastenet for kindly handed us the DL07 SMC maps. Finally, we thank the referee for the helpful and constructive comments, which have improved the manuscript.

\end{acknowledgements}

\bibliographystyle{aa}
\bibliography{Referencias.bib}

\begin{appendix}

\FloatBarrier 
\twocolumn

\begin{table*}[ht!]
\section{Measured Spectral Properties}

\caption{Spectral lines parameters summary for each SMC region.}\label{[CII] table}
\centering
\begin{tabular}{lccccccccc}
\hline \hline
 & & \cii\ & & & CO & & & H\,I & \\
Region &
$v_{\mathrm{peak}}$\tablefootmark{1} &
$\overline{\mathrm{FWHM}}$\tablefootmark{2} &
$\int I_{[\mathrm{C\,II}]}dv$\tablefootmark{3} &
$v_{\mathrm{peak}}$\tablefootmark{1} &
$\overline{\mathrm{FWHM}}$\tablefootmark{2} &
$\int I_{\mathrm{CO}}dv$\tablefootmark{3} &
$v_{\mathrm{peak}}$\tablefootmark{1} &
$\overline{\mathrm{FWHM}}$\tablefootmark{2} &
$\int I_{\mathrm{H\,I}}dv$\tablefootmark{3} \\
\hline
N83-R1 & 164.1 & 2.278 & {9.87} & 163.3 & 1.730 &  {4.638} & 182.3 & 10.39 &  {3437} \\
N83-R2 & 159.3 & 5.185 &  {12.35} & 163.7 & 2.714 &  {5.673} & 170.4 & 12.58 &  {2426} \\
N83-R3 & 161.7 & 3.707 &  {18.69} & 162.0 & 3.079 &  {14.45} & 178.0 & 16.96 &  {3701}\\
N83-R4 & 166.3 & 3.56 &  {22.05} & - & - & - & 165.2 & 16.86 &  {3502} \\
SWbarS-R5 & 124.6 & 9.240 &  {23.69} & 126.3 & 6.776 &  {15.05} & 120.96 & 22.62 &  {4544}\\
SWbarS-R6 & 124.8 & 4.158 &  {22.31} & 124.7 & 3.211 &  {4.60} & 119.4 & 25.44 &  {4306}\\
SWbDarkPeak-R7 & 140.1 & 1.249 &  {8.70} & 139.4 & 1.028 &  {2.872} & 161.6 & 21.40 &  {5315} \\
N22-R1 & 119.6 & 6.566 &  {14.81} & 121.3 & 4.723 &  {3.636} & 118.8 & 16.33 &  {4425} \\
N22-R2 & 120.9 & 4.554 &  {22.42} & 121.2 & 3.421 &  {7.60} & 131.9 & 24.14 &  {4503} \\
N22-R3 & 120.3 & 3.645 &  {18.04} & 120.3 & 2.618 &  {10.46} & 121.4 & 25.05 &  {4453} \\
SWbarN-R4 & 114.0 & 3.711 &  {48.58} & 114.9 & 3.407 &  {15.05} & 126.7 & 24.84  &  {6631} \\
SWbarN-R5 & 113.7 & 6.681 &  {22.27} & 113.3 & 5.026 &  {18.26} & 126.6 & 24.52 &  {6500} \\
SWbarN-R6 & 111.1 & 7.138 &  {23.18} & 110.3 & 4.562 &  {14.88} & 113.5 & 22.31 &  {6485} \\
\hline
\end{tabular}
\tablefoot{The integrated line intensity was computed for the rebinned spectra, while the other parameters were obtained from the original velocity resolution of each fitted spectral line.}\\ 
\tablefoottext{1}{Estimated velocity peak position of the spectral line's main Gaussian component in [km s$^{-1}$]}\\
\tablefoottext{2}{Measured line width of the spectral line's main Gaussian component in units of [km s$^{-1}$]}\\
\tablefoottext{3}{Integrated intensity [K km s$^{-1}$] of the whole line profile.}
\end{table*}

\FloatBarrier 
\twocolumn

\onecolumn
    \begin{figure*}
    \section{\cii\ 158 $\mu$m Spectral Decomposition} \label{[CII] models}
        \centering
   \includegraphics[width=18cm]{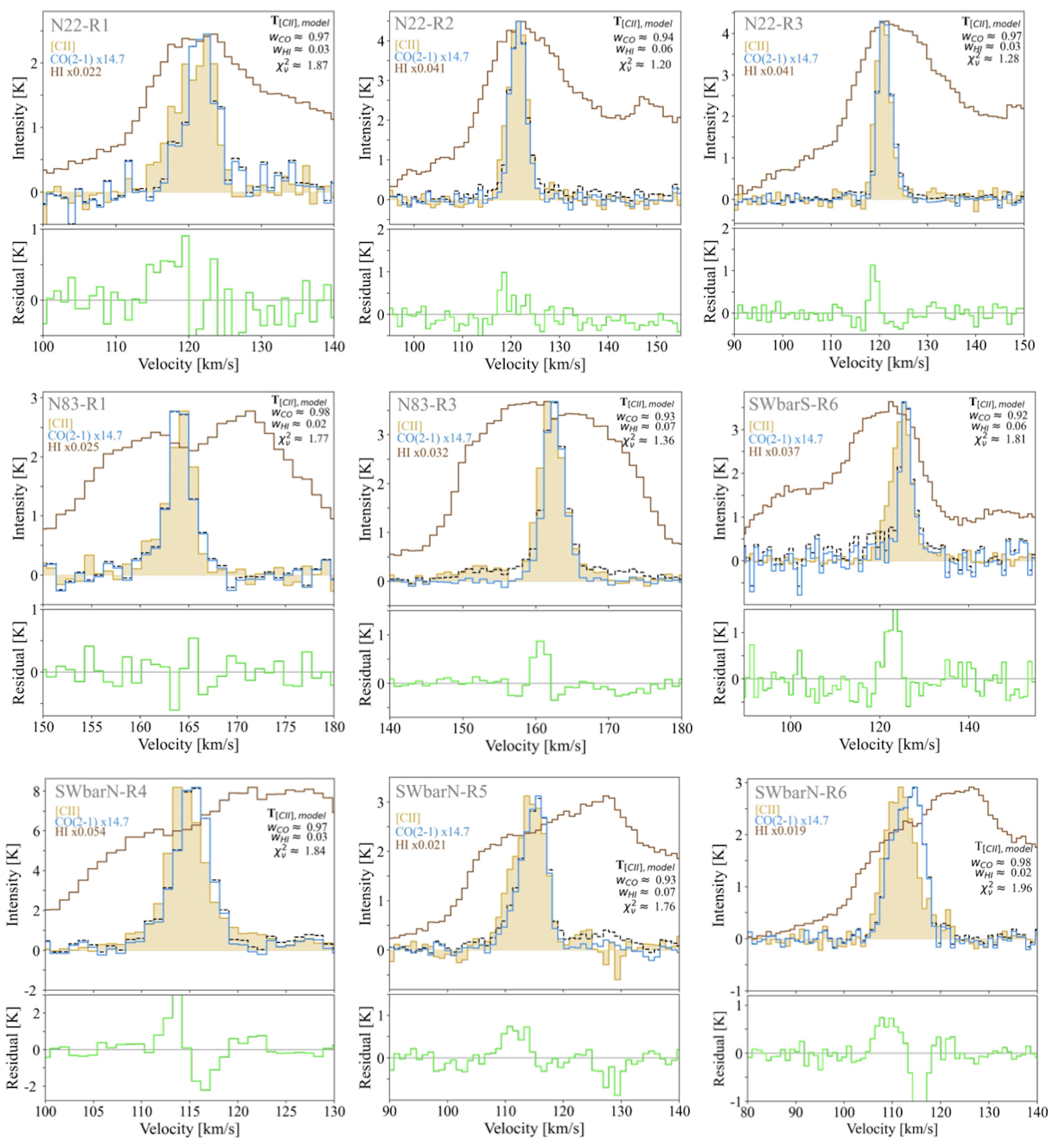}
   \caption{Rebinned \cii\ 158 $\mu$m (yellow), CO(2–1) (light blue), and \hi\ (brown) spectra for the analyzed positions in the N22, N83, SWbarN, and SWbarS regions. The CO and \hi\ spectra were scaled to the peak intensity of the corresponding \cii\ spectrum. The dashed black line shows the best-fitting \cii\ profile obtained from the linear decomposition described in Section~\ref{decomposition}. The fitted coefficients ($w_{\rm CO}$ and $w_{\text{\hi}}$) values are indicated in each panel. The lower panels display the residuals (lime) between the observed and modeled (dashed black line) \cii\ spectra. In most positions, the residual emission exceeds the spectral noise level and is preferentially concentrated around the velocity range where the CO emission peaks. This behavior suggests that the excess \cii\ emission is kinematically associated with the molecular component and is consistent with the presence of an extended reservoir of CO-faint molecular gas surrounding the denser CO-bright regions, as expected in low-metallicity environments where reduced dust shielding enhances the photodissociation of CO while allowing H$_2$ to survive \citep{Wolfire2010,Langer2014,Pineda2017}.}
    \end{figure*}

\end{appendix}
\end{document}